\documentclass[12pt]{article}

\usepackage[utf8]{inputenc}

\usepackage[margin=1in]{geometry}
\usepackage{setspace}

\usepackage{amsmath, amsthm, amssymb, amsfonts}
\usepackage{bbm}
\usepackage{bigints}

\usepackage{graphicx}
\graphicspath{{Pictures/}}
\usepackage{caption}
\usepackage{subcaption}
\usepackage{float}

\usepackage{booktabs}
\usepackage{multirow}

\usepackage{tikz}
\makeatletter
\IfFileExists{tikzlibrarybayesnet.code.tex}{\usetikzlibrary{bayesnet}}{}
\makeatother

\usepackage[toc,page]{appendix}

\usepackage{natbib}
\usepackage{url}

\usepackage[ruled]{algorithm2e}

\usepackage{xcolor}
\usepackage{enumitem}
\usepackage{enumerate}
\usepackage{authblk}
\usepackage{framed}
\usepackage[normalem]{ulem}
\usepackage{titlesec}
\usepackage{comment}
\usepackage{pdfpages}
\usepackage{epstopdf}

\usepackage[colorlinks,citecolor=blue,urlcolor=blue]{hyperref}

\def\bm{\boldsymbol}

\usepackage{amsmath}
\usepackage{amsthm}
\usepackage{amssymb}
\usepackage{graphicx}
\usepackage{color}
\usepackage{enumerate}
\usepackage{natbib}
\usepackage{url} 
\usepackage{multirow}
\usepackage{lineno}
\usepackage{makecell}
\usepackage{hyperref}

\newtheorem{theo}{\textbf{Theorem}}
\def\bvarepsilon{\bm\varepsilon}

\def\bxi{\boldsymbol \xi}

\def\f{\boldsymbol f}
\def\g{\boldsymbol g}
\def\C{\boldsymbol C}

\def\0{\boldsymbol 0}

 \title{\bf Graphical Principal Component Analysis of Multivariate Functional Time Series}
  \author{Jianbin Tan \thanks{
    Decai Liang is the co-first author.
    The authors gratefully acknowledge \textit{the National Natural Science Foundation of China (Grants No. 12231017, 12292984, and 12101332). No potential conflict of interest was reported by the author(s).}}\\
    School of Mathematics, Sun Yat-sen University\\
    Decai Liang\\
    School of Statistics and Data Science, LPMC and KLMDASR,\\
     Nankai University\\
    Yongtao Guan\\
    Shenzhen Research Institute of Big Data, School of Data Science, \\
    The Chinese University of Hong Kong, Shenzhen (CUHK-Shenzhen)\\
    and \\
    Hui Huang \\
    Center for Applied Statistics and School of Statistics, Renmin University of China}

\date{}

\begin{document}
\maketitle

\begin{abstract}
In this paper, we consider multivariate functional time series with a two-way dependence structure: serial dependence across time points and graphical interactions among multiple functions at each time point. We develop the notion of dynamic weak separability, a more general condition than those assumed in the literature, and use it to characterize the two-way structure in multivariate functional time series. Based on the proposed weak separability, we develop a unified framework for functional graphical models and dynamic principal component analysis, and further extend it to optimally reconstruct signals from contaminated functional data using graphical-level information. We investigate the asymptotic properties of the resulting estimators and illustrate the effectiveness of our proposed approach through extensive simulations. We apply our method to hourly air pollution data collected from a monitoring network in China.
\end{abstract}


\setstretch{1.5} 
\section{Introduction}
Functional data analysis (FDA) techniques have been widely used to study features of randomly sampled curves \citep{ramsay1997functional,hsing2015theoretical}. As a useful tool for dimension reduction and feature extraction, functional principal component analysis (FPCA) plays a prominent role in the analysis of functional data \citep{james2000principal,yao2005functional,ramsay1997functional}. In recent years, with the rapid development of data collection methods, multivariate functional data that possess complex temporal correlation structures have become increasingly available. Examples of such data include age-specific mortality rates \citep{gao2017multivariate,GAO2019232},
daily traffic flows \citep{chiou2014multivariate}, and social media post counts \citep{zhunetwork}.
To analyze these data, we need to account for both multivariate and temporal correlations. Challenges may arise for FPCA due to the need to simultaneously model such two-way dependencies.

To characterize multivariate dependencies, we utilize the partial correlation graph \citep{epskamp2018tutorial} for a set of random elements, in which some pairs are connected if they are partially correlated.
Graphical models provide a powerful framework to describe complex dependencies among random objects, which have been defined for both multivariate variables \citep{friedman2001elements} and multivariate time series \citep{dahlhaus2000graphical,dahlhaus2003causality}, and are called Gaussian graphical models under an additional Gaussian assumption.
Recently, \citet{qiao2019functional,qiao2020doubly} extended graphical models to characterize multivariate functional observations. However, these models are only applicable to finite-dimensional functional data. It is not trivial to explore partial correlations among infinite-dimensional functions, as the resulting covariance operator is compact and thus not invertible \citep{hsing2015theoretical}.
To solve this, \citet{zapata2019partial} introduced a separability condition to establish partial correlation graphs for infinite-dimensional curves.

Although the aforementioned works evaluate graphical models for functional data, their benefits for FPCA are rarely discussed. In addition, they may not be appropriate for temporally correlated multivariate functional data.
Multivariate functions observed over time are called multivariate functional time series (MFTS), which is a generalization of scalar time series to the multivariate functional case. 
For the FPCA of MFTS, one can employ conventional Karhunen--Lo\`eve (KL) expansions \citep{ramsay1997functional} for each individual functional time series \citep{gao2017multivariate,GAO2019232}.
However, this method may not be flexible enough to capture serial dependencies within temporally correlated functions.
Recently, \citet{hormann2015dynamic} developed a dynamic functional principal component analysis (DFPCA) approach using a frequency-domain method. With DFPCA, a univariate functional time series can be optimally reconstructed using a dynamic KL expansion. In theory, the new representation for functional time series is more general and efficient than the conventional KL expansion. 
Nonetheless, both of these methods ignore graphical interactions in MFTS, resulting in a loss of statistical efficiency for FPCA.

\begin{figure}[h]
\begin{center}
\includegraphics[scale = 0.45]{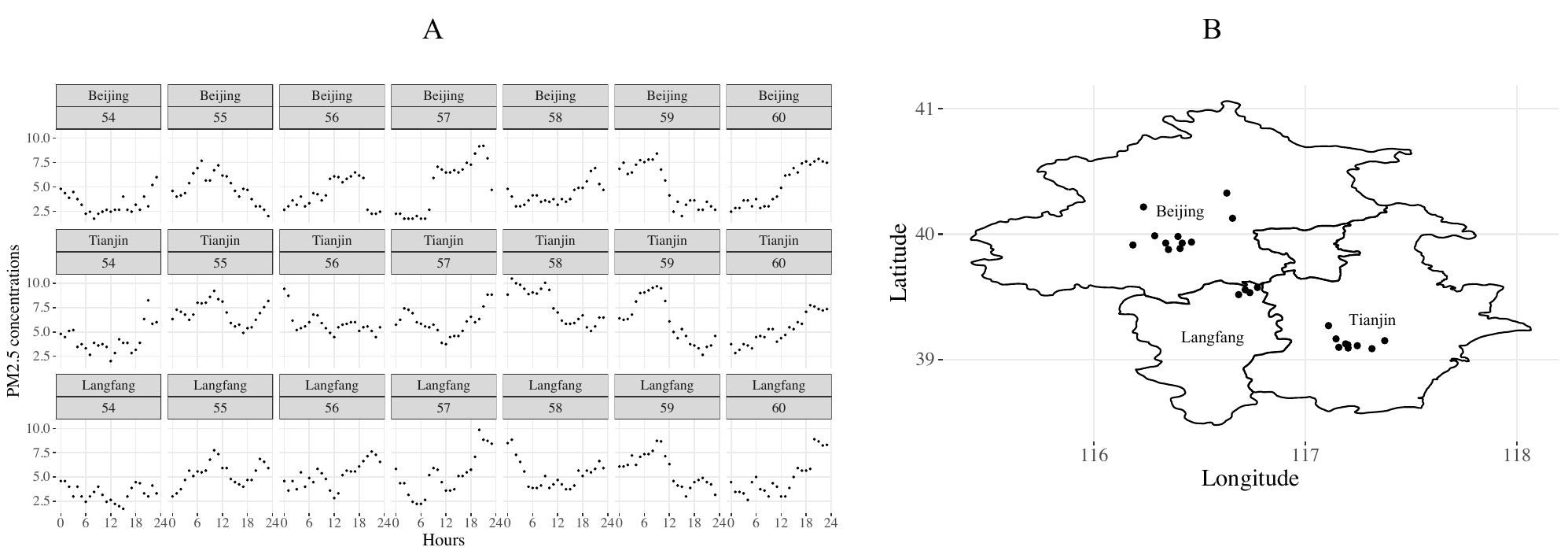}
\end{center}
\caption{\textbf{A}. Hourly readings of PM2.5 concentrations for seven days from three selected monitoring stations in Beijing, Tianjin, and Langfang, respectively. Each row is a functional time series for a specific station. \textbf{B}. Locations of 24 monitoring stations in Beijing, Tianjin, and Langfang.} \label{dat}
\end{figure}

\begin{figure}[h]
\centering
\includegraphics[scale=0.7]{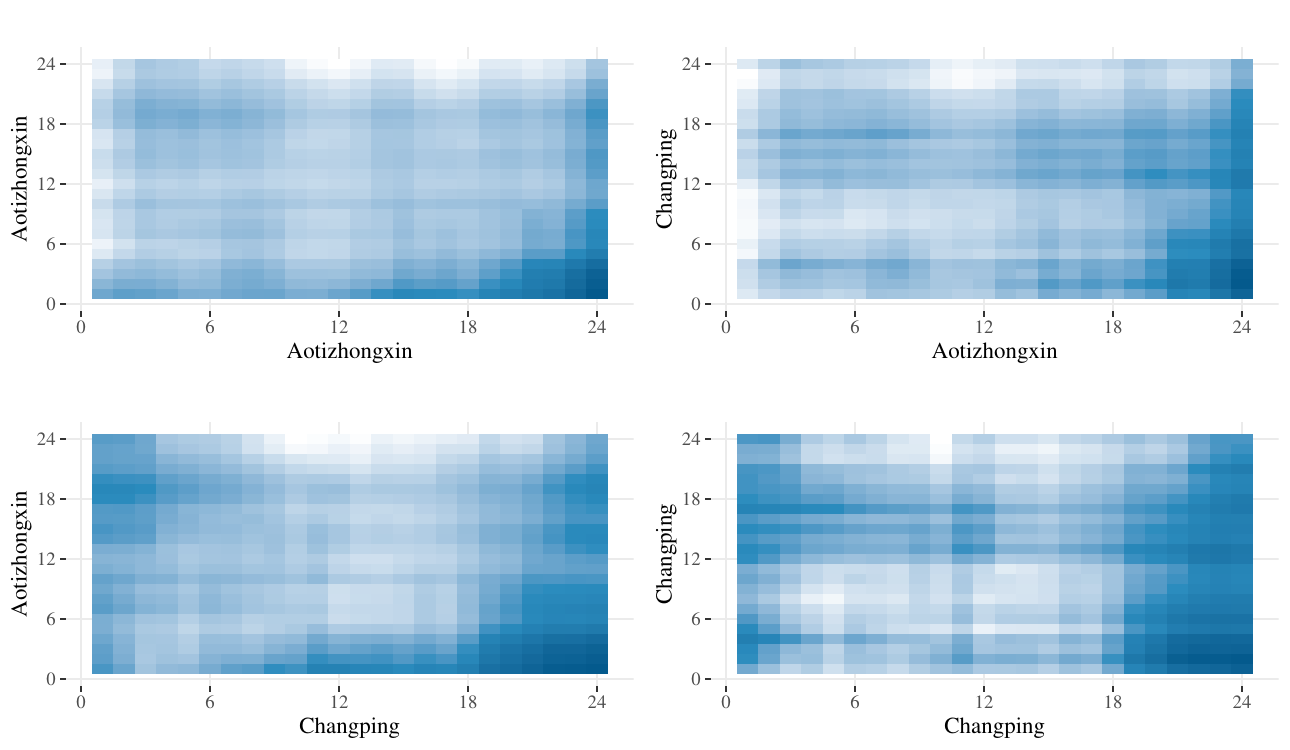}
\caption{One-day lag (hourly) autocovariance of the data from two selected stations: Changping and Aotizhongxin.}
\label{dat_co}
\end{figure}

In this article, we consider an MFTS, $\bvarepsilon_{j}(t)\in \mathbb{R}^p$, where $t\in \mathcal{T}$, with $\mathcal{T}$ being the domain of the multivariate functions from $p$ subjects, and $j$ is the index of the discrete-time unit. We assume that the MFTS has both serial dependencies over time, as indexed by $j$, and a partial correlation graph structure among the different subjects. An example of such data is illustrated in Figure~\ref{dat}, which shows seven days of hourly readings of PM2.5, a particle air pollutant, from a monitoring network in Beijing, Tianjin, and Langfang, China. On a single day, the PM2.5 concentrations from a specific station can be seen as a realization of a random function. Variations of these functions across different days and stations then show complicated spatiotemporal patterns of PM2.5 concentrations (see Figure~\ref{dat_co}). 
In this case, we treat the temporal curves across stations as an MFTS with graphical interactions.

The above MFTS data can also be treated as two-way functional data $\varepsilon_j(\bm{s}_i,t)$, where $\varepsilon_j(\bm{s}_i,t)$ is the $i^{\text{th}}$ component of $\bvarepsilon_{j}(t)$, with $\bm{s}_i$ being the location of the $i^{\text{th}}$ station in Figure~\ref{dat}\textbf{B}.
Methodological challenges in analyzing such data may arise due to their complex covariance structure at multiple levels, specifically the spatial, temporal, and daily functional levels. 
To reduce model complexity, previous works \citep{chen2017modelling,lynch2018test,bopp2022spatio} assumed simple forms of covariance structures for functional data. 
However, these assumptions are typically established based on the conventional KL expansion and fail to account for serial dependencies among MFTS. 
To address this issue, we assume a more general condition on the covariance structure of MFTS based on the dynamic KL expansion. Given this, we aim to develop an efficient FPCA for MFTS data that considers both serial dependencies and graphical interactions.

The main contributions of our work are listed as follows.
First, we propose a dynamic weak separability condition to generally characterize temporal and graphical dependencies for MFTS data in the frequency domain. 
Under this condition, we define a partial correlation graph for MFTS through a valid partial spectral density. Our framework includes the functional graphical model proposed by \citet{zapata2019partial} as a special case. 
Second, we extend DFPCA to the multivariate case. Under dynamic weak separability, we embed a graphical structure into DFPCA and provide an optimal representation for MFTS that preserves all information on the graph structure. To the best of our knowledge, this is the first theoretical framework to facilitate DFPCA for MFTS, and the resulting representation for functional data is more general than those proposed in the literature \citep{chen2017modelling,bopp2022spatio,zapata2019partial}.
Finally, we develop a novel two-step procedure to reconstruct contaminated MFTS data, 
considering both serial dependencies and graphical interactions based on the Whittle likelihood \citep{whittle1961gaussian}.

The rest of this article is organized as follows.
In Section~\ref{me}, we introduce the dynamic weak separability condition, graphical model, and DFPCA for MFTS data. 
Based on this, we propose a two-step procedure to conduct a graphical version of DFPCA (GDFPCA) for contaminated MFTS data in Section~\ref{ap}. We demonstrate the superiority of GDFPCA over existing methods through a simulation study in Section~\ref{sim}. Finally, we illustrate our method by analyzing the motivating dataset in Section~\ref{real} and present a discussion in Section~\ref{disc}. The code, data, and proofs for this article can be found in the online supplemental materials and at \href{https://github.com/Jianbin-Tan/GFPCA}{https://github.com/Jianbin-Tan/GFPCA}.

\section{A Unified Framework for Graphical Models and FPCA}\label{me}
Define $\langle \f,\g\rangle_{p}:=\int_{x_1\in \mathcal{T}}\cdots \int_{x_d\in \mathcal{T}}\big\{\f(x_1,\cdots,x_d)\big\}^*\g(x_1,\cdots,x_d)\  \mathrm{d}x_1\cdots \mathrm{d}x_d$ and let $||\f||_{p}:=\sqrt{\langle \f,\f\rangle_{p}}$ as the inner product and $L^2$-norm, respectively, $\forall \f(\cdot),\g(\cdot)\in L^2(\mathcal{T}^d,\mathbb{C}^p)$, where 
$\mathcal{T}^d$ is a compact subset contained in a $d$-dimensional real Euclidean space, $\mathbb{C}^p$ is the $p$-dimensional complex Euclidean space, $\{\cdot\}^*$ is the transposed conjugate operation upon a complex-valued matrix, and 
$L^2(\mathcal{T}^d,\mathbb{C}^p)$ is the collection of all square-integrable functions mapping from $\mathcal{T}^d$ to $\mathbb{C}^p$. The notation $\langle \cdot,\cdot \rangle_{p}$ and $||\cdot||_{p}$ will be simplified as $\langle \cdot,\cdot \rangle$ and $||\cdot||$ if $p=1$. Besides, $|\cdot|$ and $\overline{(\cdot)}$ are the length and the conjugate of a complex number or vector, respectively, and $\operatorname{tr}(\cdot)$ is the trace of a matrix. Moreover, $\mathcal{H}(\mathcal{T}^d,\mathbb{R}^p)$ contains a collection of functions $\{\bm{h}_g(\cdot);g\in \mathbb{Z}\}$ s.t. $\bm{h}_g(\cdot)\in L^2(\mathcal{T}^d,\mathbb{R}^p)$, $\forall$ $g\in \mathbb{Z}$, and there exists a function ${\bm{H}}(\cdot|\theta)\in L^2(\mathcal{T}^d,\mathbb{C}^p)$ indexed by $\theta$ s.t. $\int_{-\pi}^{\pi}\left\|{\bm{H}}(\cdot|\theta)\right\|_p^2\ \mathrm{d}\theta<\infty$ and $\lim_{G\rightarrow\infty}\int_{-\pi}^{\pi}\left\|{\bm{H}}(\cdot|\theta)-\frac{1}{2\pi}\sum_{|g|\leq G}\bm{h}_g(\cdot)\exp(\operatorname{i}g\theta)\right\|_p^2\ \mathrm{d}\theta= 0$, where $\operatorname{i}$ is the imaginary unit. A sufficient condition for $\{\bm{h}_g(\cdot);g\in \mathbb{Z}\}\in \mathcal{H}(\mathcal{T}^d,\mathbb{R}^p)$ is $\sum_{g\in \mathbb{Z}}\left\|\bm{h}_g\right\|_p<\infty$, where we have ${\bm{H}}(\cdot|\theta)=\frac{1}{2\pi}\sum_{g\in\mathbb{Z}}\bm{h}_g(\cdot)\exp(\operatorname{i}g\theta)$ and $\bm{h}_g(\cdot)=\int_{-\pi}^{\pi}\bm{H}(\cdot|\theta)\exp(-\operatorname{i}g\theta)\ \mathrm{d}\theta$.

For each $j$, we assume that $\bvarepsilon_j(t)\in \mathbb{R}^p$, is a zero-mean multivariate process on $t\in \mathcal{T}$ s.t. $\mathbb{E}||\bvarepsilon_j||_p^2<\infty$. For simplicity, $\mathcal{T}$ is $[0,1]$ in our study. The $i^{th}$ component of $\bvarepsilon_j(t)$ is denoted as $\varepsilon_{ij}(t)$ for $i\in V:=\left\{1,\cdots,p\right\}$. 
For ease of presentation, $\bvarepsilon_{j}(t)$ and $\varepsilon_{ij}(t)$ are abbreviated as $\bvarepsilon_{j}$ and $\varepsilon_{ij}$, respectively. 
We assume that $\bvarepsilon_j$ possesses temporal dependencies for different $j$, and the MFTS $\{\bvarepsilon_{j};j\in \mathbb{Z}\}$ is weakly stationary, i.e., $\forall j_1,j_2\in \mathbb{Z}$ and $t,s\in [0,1]$,
\begin{equation}\label{wst}
\mathbb{E}\bvarepsilon_{j_1}(t)\bvarepsilon_{j_2}(s)^T=\C_{g}(t,s)\in \mathbb{R}^{p\times p},
\end{equation}
where $g=j_1-j_2$.
Let $C_{i_1,i_2,g}(t,s)$ denote the $(i_1,i_2)^{th}$ element of $\C_{g}(t,s)$, and assume $\sup_{i_1,i_2\in V}\sum_{g\in \mathbb{Z}}||C_{i_1,i_2,g}||<\infty$. We define the spectral density kernel as 
\begin{equation}\label{ft1}
\f(t,s|\theta):=\frac{1}{2\pi}\sum_{g\in \mathbb{Z}}\C_{g}(t,s) \exp(\operatorname{i}g\theta),\ \theta \in [-\pi, \pi],\ t,s\in[0,1].
\end{equation} 
Conversely, $\C_{g}(t,s)$ can be represented by the inverse Fourier transform of $\f(t,s|\theta)$, i.e., 
$\C_{g}(t,s)=\int_{-\pi}^{\pi}\f(t,s|\theta)\exp(-\operatorname{i}g\theta)\ \mathrm{d}\theta,\ g \in \mathbb{Z},\ t,s\in[0,1].$
Similarly, we denote the $(i_1,i_2)^{th}$ element of $\f(t,s|\theta)$ as $f_{i_1,i_2}(t,s|\theta)$. 

\subsection{Dynamic Weak Separability}\label{GWSC}

We assume that the spectral density kernel $\f(t,s|\theta)$ is continuous for all $t,s\in [0,1]$ and $\theta\in [-\pi,\pi]$. Then by the general Mercer's theorem \citep{de2013extension}, $\f(t,s|\theta)$, for each $\theta\in [-\pi,\pi]$, admits the following spectral decomposition:
\begin{equation}
    \label{spd}
\f(t,s|\theta)=\sum_{k=1}^{\infty}\omega_{k}(\theta){\bm{\delta}_{k}(s|\theta)} \left\{\bm{\delta}_{k}(t|\theta)\right\}^*,
\end{equation}
where $\omega_{1}(\theta)\geq \omega_{2}(\theta)\geq \cdots\geq 0$ are the eigenvalues and $\bm{\delta}_{k}(\cdot|\theta)$, $k\geq 1$, are the eigenfunctions in $L^2([0,1],\mathbb{C}^p)$. We say that $\f(t,s|\theta)$ is weakly separable if
\begin{equation}\label{ws}
\f(t,s|\theta)=\sum_{k=1}^{\infty} \bm{\eta}_{k}(\theta) \overline{\psi_{k}(t|\theta)} {\psi_{k}(s|\theta)},
\end{equation}
where $\psi_{k}(\cdot|\theta)$, $k\geq 1$, are orthonormal functions in $L^2([0,1],\mathbb{C})$,
and the eigen-matrix $\bm{\eta}_{k}(\theta) =\left\{\eta_{i_1,i_2,k}(\theta)\right\}_{1\leq i_1,i_2\leq p}$ is non-negative definite for each $k$ and $\theta$ satisfying $\operatorname{tr}\{\bm{\eta_1}(\theta)\}\geq \operatorname{tr}\{\bm{\eta_2}(\theta)\}\geq \cdots\geq 0$ and $\sum_{k=1}^\infty\int_{-\pi}^{\pi}\operatorname{tr}\{\bm{\eta_k}(\theta)\}\ \mathrm{d}\theta<\infty$. Under the weak separability, the scalar-valued eigenvalue and vector-valued eigenfunction, i.e., $\omega_k(\theta)$ and $\bm{\delta}_k(\cdot|\theta)$ in \eqref{spd}, become the eigen-matrix and scalar-valued eigenfunction, i.e., $\bm{\eta}_k(\theta)$ and $\psi_k(\cdot|\theta)$ in \eqref{ws}, for each $\theta$. This modification can greatly facilitate our investigation of the graphical structure and FPCA of the MFTS in Section \ref{GDFPCA}. Note that \eqref{ws} holds if each $\bm{\delta}_{k}(\cdot|\theta)$ in \eqref{spd} can be factorized as the product of a scalar function and a $p$-dimensional vector for each $\theta$; see Lemma 2 in Supplementary Materials for more details.

A concept that is closely related to our proposed weak separability of the spectral density kernel is the so-called weak separability of the covariance function.
We say that $\C_{g}(t,s)$ is weakly separable if 
\begin{equation}\label{wss}
\C_{g}(t,s)=\sum_{k=1}^{\infty} \bm{\vartheta}_{k,g}\ {\varphi_{k}(t)} {\varphi_{k}(s)},
\end{equation} 
given some matrices $\bm{\vartheta}_{k,g}$s and orthonormal eigenfunctions $\varphi_{k}(\cdot)$s. The weak separability above was introduced in \citet{liang2021test} for $p=1$ and we extended it to $p>1$. When $\C_{g}(t,s)=0$ for $g \neq 0$ and $t,s\in[0,1]$, \eqref{wss} was also called the partial separability in \citet{zapata2019partial} and can degenerate to the weak separability in \citet{lynch2018test} under other conditions.
To make the terminologies consistent, we use weak separability to refer to the weak separability in terms of covariance functions, i.e., \eqref{wss}. Besides, we call the weak separability \eqref{ws} the dynamic weak separability, since it is defined on the frequency domain.
Note that \eqref{wss} trivially implies \eqref{ws} due to \eqref{ft1}. It can be shown that the converse is true iff $\psi_{k}(\cdot|\theta)$ can be separated as $\psi_{k}(\cdot|\theta)=\gamma_k(\theta)\varphi_{k}(\cdot)$, $\forall$ $\theta\in [-\pi,\pi]$, with $\gamma_k(\theta)$ being some multiplicative factors on the complex unit circle. Thus, the dynamic weak separability \eqref{ws} is a more general condition. 

\subsection{Graphical Model for MFTS}\label{gra}
Let $\bvarepsilon_{V_1,j}(t)$ be the sub-vector of $\bvarepsilon_j(t)$ containing the coordinates in $V_1\subset V$. We can then define $\bvarepsilon_{V_1,\cdot}:=\{\bvarepsilon_{V_1,j};j\in \mathbb{Z}\}$ and for a special case $V_1=\{i\}$, we have $\bvarepsilon_{i,\cdot}:=\{\varepsilon_{ij};j\in \mathbb{Z}\}$. We say that $\bvarepsilon_{i_1,\cdot}$ and $\bvarepsilon_{i_2,\cdot}$ are uncorrelated (denoted as $\bvarepsilon_{i_1,\cdot}\perp \bvarepsilon_{i_2,\cdot}$) iff 
$C_{i_1,i_2,g}(t,s)=0$ for all $g\in \mathbb{Z}$ and $t,s\in [0,1]$, or equivalently, $f_{i_1,i_2}(t,s|\theta)=0$ for all $t,s\in [0,1]$ and $\theta\in [-\pi,\pi]$.

Let $V_{-\{ i_1,i_2\}}:=V\backslash\{i_1,i_2\}$. In this subsection, we first define partial uncorrelatedness between $\bvarepsilon_{i_1,\cdot}$ and $\bvarepsilon_{i_2,\cdot}$ given $\bvarepsilon_{V_{-\{ i_1,i_2\}},\cdot}$, denoted as $\bvarepsilon_{i_1,\cdot}\perp\bvarepsilon_{i_2,\cdot}| \cdot$. Accordingly, we define a graph $(V,E)$ based on $\bvarepsilon_{V,\cdot}$, where the edge set $E\subset V^2$ contains $\big\{(i,i);i\in V\big\}$ and all pairs of distinct indices $(i_1,i_2)$ such that $\bvarepsilon_{i_1,\cdot}$ and $\bvarepsilon_{i_2,\cdot}$ are partially correlated. 

To formally introduce the definition, we follow the scheme that defines graphical models for multivariate time series \citep{dahlhaus2000graphical}. Specifically, we at first remove the linear effect of $\bvarepsilon_{V_{-\{ i_1,i_2\}},\cdot}$ from $\bvarepsilon_{i_1,\cdot}$ and $\bvarepsilon_{i_2,\cdot}$. Take $\bvarepsilon_{i_1,\cdot}$ as an example, we define ${\varepsilon}^{\text{r}}_{i_1j}:=\varepsilon_{i_1j}-{\varepsilon}^{\text{p}}_{i_1j}$, where
\begin{equation}{\varepsilon}^{\text{p}}_{i_1j}=\arg\min_{{\varepsilon}^{\prime}_{i_1j}\in L_{ i_1,i_2}^2(\mathcal{T},\mathbb{R})}\int_0^1\mathbb{E}\big\{\varepsilon_{i_1j}(t)-{\varepsilon}^{\prime}_{i_1j}(t)\big\}^2\ \mathrm{d}t,
\end{equation}
among $L_{i_1,i_2}^2(\mathcal{T},\mathbb{R})$ is the closure of all linear predictors on $\bvarepsilon_{V_{-\{ i_1,i_2\}},\cdot}$ in the sense of $L^2$-norm. In other word, $\forall$ ${\varepsilon}^{\prime}_{i_1j}\in L_{i_1,i_2}^2(\mathcal{T},\mathbb{R})$ and $\forall\delta>0$, there exists $\{\bm{a}_{i_1,g}(\cdot,\cdot);g\in\mathbb{Z}\}\in \mathcal{H}([0,1]^2,\mathbb{R}^{p-2})$ s.t. $\int_0^1\mathbb{E}\left\{{\varepsilon}^{\prime}_{i_1j}(t)-\sum_{g\in\mathbb{Z}}\langle\bm{a}_{i_1,g}(t,\cdot),\bvarepsilon_{V_{-\{ i_1,i_2\}},j+g}\rangle_{p-2}\right\}^2\ \mathrm{d}t<\delta$.
Accordingly, $\bvarepsilon^{\text{r}}_{i_1,\cdot}$ and $\bvarepsilon^{\text{r}}_{i_2,\cdot}$ can be viewed as the residual functional time series obtained by regressing $\bvarepsilon_{i_1,\cdot}$ and $\bvarepsilon_{i_2,\cdot}$ on $\bvarepsilon_{V_{-\{ i_1,i_2\}},\cdot}$. Note that both $\bvarepsilon^{\text{r}}_{i_1,\cdot}$ and $\bvarepsilon^{\text{r}}_{i_2,\cdot}$ are weakly stationary processes. Denote ${C}_{i_1,i_2|\cdot,g}\left(t,s\right)$ as the cross-covariance between $\bvarepsilon^{\text{r}}_{i_1,\cdot}$ and $\bvarepsilon^{\text{r}}_{i_2,\cdot}$. We say that
\begin{equation}
\bvarepsilon_{i_1,\cdot}\perp\bvarepsilon_{i_2,\cdot}| \cdot\ \text{if}\ \bvarepsilon^{\text{r}}_{i_1,\cdot}\perp \bvarepsilon^{\text{r}}_{i_2,\cdot},\label{cd}
\end{equation}
that is, ${C}_{i_1,i_2|\cdot,g}\left(t,s\right)=0$, $\forall g\in \mathbb{Z}$ and $t,s\in [0,1]$, or the associated partial spectral density ${f}_{i_1,i_2|\cdot}(t,s|\theta)=0$, $\forall$ $t,s\in [0,1]$ and $\theta\in [-\pi,\pi]$. The defined graph $(V,E)$ via relation (\ref{cd}) is called a partial correlation graph for the MFTS.
\begin{theo}\label{th2}
Under the dynamic weak separability (\ref{ws}), we assume that for all $\theta\in[-\pi,\pi]$ and $k\geq 1$, $\bm{\eta}_k(\theta)$ is a nonsingular matrix. Then for $i_1\neq i_2$,
\begin{equation}\label{fco}
 {f}_{i_1,i_2|\cdot}(t,s|\theta)=\sum_{k=1}^{\infty} \sigma_{i_1,i_2,k}(\theta) \overline{\psi_{k}(t|\theta)} {\psi_{k}(s|\theta)},
 \end{equation}
with $\sigma_{i_1,i_2,k}(\theta)=-\frac{[\bm{\Phi}_k(\theta)]_{i_1,i_2}}{[\bm{\Phi}_k(\theta)]_{i_1,i_1}[\bm{\Phi}_k(\theta)]_{i_2,i_2}-[\bm{\Phi}_k(\theta)]_{i_1,i_2}[\bm{\Phi}_k(\theta)]_{i_2,i_1}}$, where $\bm{\Phi}_k(\theta)=\big\{\bm{\eta}_k(\theta)\big\}^{-1}$ and $[\cdot]_{V_1,V_2}$ is the operation to extract subsets $V_1,V_2$ of rows and columns of a matrix. Accordingly,
\begin{eqnarray*}
 (i_1,i_2)\notin E\ \text{iff}\ [\bm{\Phi}_k(\theta)]_{i_1,i_2}=0,\ \forall \theta\in[-\pi,\pi]\ \text{and}\ k\geq 1.
 \end{eqnarray*}
\end{theo}

Theorem \ref{th2} gives a form of the partial spectral density kernel for an infinite-dimensional MFTS under the dynamic weak separability. It also establishes a connection between the partial correlation graph $(V,E)$ and $\bm{\Phi}_k(\theta)$; see Part A.2 in Supplementary Materials for the detailed proof.
It's worth noting that the graphical model in \citet{zapata2019partial} is our special case when $\bm{C}_g(t,s)=0$ for all $t,s\in [0,1]$ and $g\neq 0$. 
Therefore, our framework is more general in defining a graphical model for MFTS. 

\subsection{Dynamic Weakly-Separable KL Expansion}\label{dfpca&ws}

In this subsection, we extend the DFPCA \citep{hormann2015dynamic} to the MFTS case under the dynamic weak separability \eqref{ws}. Let $\varepsilon_j$ and $\delta_k(\cdot|\theta)$ be the univariate counterparts to $\bm{\varepsilon}_j$ and $\bm{\delta}_k(\cdot|\theta)$, respectively. Then, the univariate dynamic KL expansion \citep{hormann2015dynamic} for $\varepsilon_{j}$ can be written as 
\begin{equation}\label{ukl}
    \varepsilon_{j}(t)=\sum_{k\geq 1}\sum_{l\in \mathbb{Z}}{\phi}_{kl}(t){\xi}_{(j+l)k},
\end{equation}
where ${\phi}_{kl}(t)=\frac{1}{2\pi}\int_{-\pi}^{\pi}{\delta}_{k}(t|\theta)\exp(-\operatorname{i}l\theta) \ \mathrm{d}\theta$ and ${\xi}_{jk}=\sum_{l\in\mathbb{Z}}\langle {\varepsilon}_{j-l}, {\phi}_{kl}\rangle$. Specifically, $\{\phi_{kl}(\cdot);l\in \mathbb{Z}\}$ are called the functional filters \citep{hormann2015dynamic} for $\{ \varepsilon_{j};j\in \mathbb{Z}\}$ and ${\xi}_{jk}$ is the projected score on the functional filters. Although \eqref{ukl} can be applied to each individual series of the MFTS separately, such an approach ignores the interdependence among the different individual series and could therefore result in a significant loss of efficiency. Moreover, it's difficult to compare the scores extracted from the different individual series, because their associated {functional filters} may be unrelated.

Based on \eqref{ukl}, we can similarly define the dynamic multivariate KL expansion of $\bm{\varepsilon}_{j}$ as
\begin{equation}\label{mkl}
    \bm{\varepsilon}_{j}(t):=\sum_{k\geq 1}\sum_{l\in \mathbb{Z}}\bm{\phi}_{kl}(t){\xi}_{(j+l)k},
\end{equation}
where $\bm{\phi}_{kl}(t)=\frac{1}{2\pi}\int_{-\pi}^{\pi}\bm{\delta}_{k}(t|\theta)\exp(-\operatorname{i}l\theta) \ \mathrm{d}\theta$ and ${\xi}_{jk}=\sum_{l\in\mathbb{Z}}\langle \bm{\varepsilon}_{j-l}, \bm{\phi}_{kl}\rangle_p$. 
Intuitively, this representation is constructed by joining the MFTS into a univariate one, i.e., connecting the starting points and ending points of the trajectories $\{\varepsilon_{ij};i\in V\}$ to form one single trajectory for each $j$. After that, we can reconstruct $\bm{\varepsilon}_{j}$ by applying \eqref{ukl} to the jointed trajectory.
From this perspective, the dynamic multivariate KL expansion can borrow strength across the individual series to extract the common scalar scores ${\xi}_{jk}$. However, because the scores are common for all individual series, they do not contain information on the potential graphical structure of the MFTS.

In view of the aforementioned issues with using \eqref{ukl} and \eqref{mkl}, we develop a dynamic multivariate KL expansion under the dynamic weak separability \eqref{ws}. Theorem \ref{es_exa} below provides the theoretical justification for our approach.

\begin{theo}\label{es_exa}
Under the weak stationarity \eqref{wst}, the following statements are equivalent:\\
(a) The dynamic weak separability \eqref{ws} is satisfied with $\eta_k(\theta)$s being nonsingular.\\
(b) There exist orthonormal basis functions of $L^2([0,1],\mathbb{C})$: $\{\psi_k(\cdot|\theta);k\geq 1\}$, $\forall\theta\in [-\pi,\pi]$, such that the scores ${\xi}_{ijk}$ are uncorrelated for different $k$, where 
\begin{eqnarray}
    {\xi}_{ijk}&=&\sum_{l\in\mathbb{Z}}\langle \varepsilon_{i(j-l)}, {\phi}_{kl}\rangle\label{dfpca2}\\
\text{with}\quad \phi_{kl}(t)&=&\frac{1}{2\pi}\int_{-\pi}^{\pi}\psi_{k}(t|\theta)\exp(-\operatorname{i}l\theta) \ \mathrm{d}\theta.\label{dfpca}
\end{eqnarray}
Define $\bxi_{\cdot,jk}=\left(\xi_{1jk},\cdots,\xi_{pjk}\right)^T$, we have that $\{\bxi_{\cdot,jk};j\in \mathbb{Z}\}$ is a weakly stationary multivariate time series with the spectral density matrix $\bm{\eta}_{k}(\theta)$. 
Furthermore, the dynamic multivariate KL expansion becomes
\begin{eqnarray}\label{newkl}
    {\bvarepsilon}_{j}(t)=\sum_{k\geq 1}\sum_{l\in \mathbb{Z}}{\phi}_{kl}(t){\bxi}_{\cdot,(j+l)k}.
\end{eqnarray}
\end{theo}

The proof of Theorem \ref{es_exa} is given in Part A.3 in Supplementary Materials.
Essentially, the dynamic weak separability \eqref{ws} suggests that the dynamic multivariate KL expansion reduces to its univariate version with a common set of functional filters, i.e., $\{\phi_{kl}(\cdot);l\in \mathbb{Z}\}$ given in \eqref{dfpca}. A similar result was also reported in \citet{zapata2019partial} under the weak separability of covariance functions.
We can prove that the functional filters satisfy 
\begin{equation}\label{Prop_1}
    \sum_{l\in\mathbb{Z}}\big\|\phi_{kl}\big\|^{2}=\frac{1}{2\pi}\int_{-\pi}^{\pi}\big\|\psi_k(\cdot|\theta)\big\|^2\ \mathrm{d}\theta=1, \ \forall k\geq 1.
\end{equation}

There are several advantages to using our proposed expansion \eqref{newkl}. Firstly, because the functional filters in \eqref{newkl} are common across the individual series, they can be estimated by using information from the entire MFTS rather than a single series. This can help significantly improve the statistical efficiency compared to the univariate expansion \eqref{ukl}, especially when the temporal information in the MFTS is limited. Secondly, again because the functional filters are common, the scores $\{\bxi_{\cdot, jk};j\in \mathbb{Z}\}$ are naturally aligned across different individual series for each $k$, which can in turn enhance interpretability of the estimation. Thirdly, we show in the next subsection that under the dynamic weak separability (\ref{ws}), the scores preserve all information on the graph structure of the MFTS, which is an advantage over the dynamic multivariate KL expansion \eqref{mkl}. 

\subsection{Graphical Functional Principal Component Analysis}\label{GDFPCA}
Given a positive integer $K$, consider the truncated dynamic weakly-separable KL expansion
\begin{equation}\label{fgts_model}
  \bvarepsilon_{j}^K(t):=\sum_{k\leq K}\sum_{l\in \mathcal{Z}}\phi_{kl}(t)\bxi_{\cdot,(j+l)k}.
\end{equation}
In this subsection, we will show that $\bm{\varepsilon}_{j}^{K}$ serves as an optimal representation of $\bm{\varepsilon}_{j}$ under the dynamic weak separability. 
Note that $\psi_{k}(\cdot|\theta)$ in (\ref{ws}) is unique up to some multiplicative factor on the complex unit circle. 
Therefore, $\phi_{kl}(\cdot)$ and $\bxi_{\cdot,jk}$ cannot be uniquely identified.
Nonetheless, we can show that $\sum_{l\in \mathcal{Z}}\phi_{kl}(t)\bxi_{\cdot,(j+l)k}$
is unique for each $j$, $k$ and $t$; see the remark in Part A.3 of Supplementary Materials for the detailed derivation. 

\begin{theo}\label{p2}
For any arbitrary $\{\tilde{\phi}_{kl}(\cdot);l\in \mathbb{Z}\}\in\mathcal{H}([0,1],\mathbb{R})$ for $k\leq K$, define $\tilde{\varepsilon}_{ij}^K(t):= \sum_{k\leq K}\sum_{l\in \mathcal{Z}}\tilde\phi_{kl}(t)\tilde{\xi}_{i(j+l)k}$ with $\tilde{\xi}_{ijk}=\sum_{l\in\mathbb{Z}}\langle \varepsilon_{i(j-l)}, \tilde{\phi}_{kl}\rangle$ and $\tilde{\bvarepsilon}_j^K(t):=(\tilde{\varepsilon}_{1j}^K(t),\cdots, \tilde{\varepsilon}_{pj}^K(t))^T$. 
Then, under the dynamic weak separability (\ref{ws}),
\begin{eqnarray*}
\mathbb{E}||\bvarepsilon_{j}-\bvarepsilon_{j}^K||_{p}^2=\sum_{k> K}\int_{-\pi}^{\pi}\operatorname{tr}\{\bm{\eta_k}(\theta)\}\ \mathrm{d}\theta\leq \mathbb{E}||\bvarepsilon_{j}-\tilde{\bvarepsilon}_{j}^K||_{p}^2,
\end{eqnarray*}
where the equality holds if $\{\tilde{\phi}_{kl}(\cdot);l\in \mathbb{Z}\}$ are constructed using (\ref{dfpca}) for all $k\leq K$.
\end{theo}

The proof of Theorem \ref{p2} is given in Part A.4 in Supplementary Materials. 
This theorem shows that $\bvarepsilon_{j}^K$ is the optimal $K$-truncated approximation of $\bvarepsilon_{j}$ in the sense of the $L^2$-norm.

There is a connection between the scores $\{\bxi_{\cdot,jk};j\in \mathbb{Z}\}$ and the graphical model mentioned in Section \ref{gra} under the dynamic weak separability (\ref{ws}). Firstly, recall that $\{\bxi_{\cdot, jk};j\in \mathbb{Z}\}$ is a weakly stationary multivariate time series with the spectral density matrix $\bm{\eta}_{k}(\theta)$, and thus we can show that $\sigma_{i_1,i_2,k}(\theta)$ in (\ref{fco}) is the partial cross-spectrum of $\{\bxi_{\cdot, jk};j\in \mathbb{Z}\}$ \citep{brillinger2001time}; see the equation (5) in Supplementary Materials for the deviation. By the partial correlation graph defined for weakly stationary multivariate time series \citep{dahlhaus2000graphical}, $\{\xi_{i_1jk};j\in \mathbb{Z}\}$ and $\{\xi_{i_2jk};j\in \mathbb{Z}\}$ are partially uncorrelated iff $\sigma_{i_1,i_2,k}(\theta)=0,\ \forall$ $\theta\in [-\pi,\pi]$. Let $(V,E)$ and $(V,E_k)$ be the partial correlation graphs of $\{\bm{\varepsilon}_j;j\in \mathbb{Z}\}$ and $\{\bxi_{\cdot,jk};j\in \mathbb{Z}\}$, respectively. It follows immediately from Theorem \ref{th2} that
\begin{equation}\label{uni_gra}
    E=\cup_{k=1}^{\infty} E_k.
\end{equation}
In other words, by using a common set of functional filters in MFTS, the extracted scores preserves all information on the graph structure of the original MFTS. This forms the basis for estimating functional filters and scores by utilizing graphical-level information, as demonstrated in the next section. Therefore, we call this kind of FPCA the graphical DFPCA (GDFPCA).

If we further assume the stronger separability condition \eqref{wss} on covariance functions, then the GDFPCA would have a simpler form; see the next theorem.

\begin{theo}\label{th33}
Under the dynamic weak separability (\ref{ws}), the following three statements are equivalent:\\
(a) The weak separability \eqref{wss} is achieved.\\
(b) There exist orthonormal basis functions of $L^2([0,1],\mathbb{R})$: $\{\varphi_k(\cdot);k\geq 1\}$, such that $\phi_{kl}(t)=c_l\varphi_{k}(t)$, $\forall k\geq 1,\ l\in \mathbb{Z}$, and $t\in [0,1]$, given some $c_l\in \mathbb{R}$.\\
(c) The optimal representation (\ref{fgts_model}) degenerates to the static version, i.e.,
\begin{equation}\label{Sta_wes}
    \varepsilon_{ij}^K(t)=\sum_{k=1}^{K}\varphi_{k}(t)\xi_{ijk}\  \text{with}\  \xi_{ijk}=\langle\varepsilon_{ij},\varphi_k\rangle.
\end{equation}
\end{theo}

The proof of Theorem \ref{th33} is given in Part A.5 in Supplementary Materials. 
Theorem \ref{th33} shows that the form of weak separability affects the optimal representation of the MFTS. Under the weak separability \eqref{wss}, the GDFPCA would degenerate to its static version as proposed in \citet{chen2017modelling,bopp2022spatio} and \citet{zapata2019partial}.
Note that the scores in \eqref{Sta_wes} also preserve all information on the graph structure of the MFTS. 
We similarly call this type of FPCA the graphical static FPCA (GSFPCA).

\section{Graphical DFPCA for Contaminated Data}\label{ap}
In this section, we use our proposed GDFPCA to reconstruct signals from contaminated MFTS data with graphical interactions. 
Let $X_{ij}(t)$ be some random functions on $t\in[0,1]$.
In practice, $X_{ij}(\cdot)$ may be observed only at some given time points and with measurement errors. We therefore assume that
\begin{equation}\label{model_dat}
Y_{ijz}=X_{ij}(t_z)+\tau_{ijz}=\mu_{i}(t_z)+\varepsilon_{ij}(t_z)+\tau_{ijz},
\end{equation}
for $i\in V,\ j=1,\cdots, J\ \text{and}\ z=1,\cdots, Z$,
where $t_z$ is a fixed time point in $[0,1]$, $\mu_i(\cdot)$ is a fixed mean function, $\varepsilon_{ij}$ is zero-mean Gaussian process, and $\tau_{ijz}$ is a zero-mean Gaussian white noise with finite variance $\sigma_i^2$ for each $i$. 
Recall $\bm{\bvarepsilon}_{j}:=\left(\varepsilon_{1j},\cdots,\varepsilon_{pj}\right)^T$, we assume $\{\bvarepsilon_j;j\in \mathbb{Z}\}$ satisfying the dynamic weak separability \eqref{ws} and possessing a partial correlation graph $(V,E)$ as defined in Section \ref{gra}.

Our main objective is to reconstruct $X_{ij}(\cdot)$ for $i\in V$ and $j=1,\cdots,J$ from the observed contaminated data $\bm{Y}:=\{Y_{ijz};i\in V,\ j=1,\cdots, J,\ z=1,\cdots, Z\}$. For this, we propose a two-step procedure based on the GDFPCA framework. Specifically, we first estimate the mean functions, and the functional filters and eigen-matrices of $\{\bvarepsilon_j;j\in \mathbb{Z}\}$, and then predict the scores by a conditional expectation estimation. 

\subsection{First-Step Estimation}\label{est of ef}
Abbreviate the kernels ${f}_{i_1,i_2}(\cdot,\cdot|\theta)$ and $\hat{f}_{i_1,i_2}(\cdot,\cdot|\theta)$ as ${f}_{i_1,i_2,\theta}$ and $\hat{f}_{i_1,i_2,\theta}$, respectively.
We first consider the case of fully observed processes $\{\bvarepsilon_j;j=1,\cdots,J\}$. 
For this case, the estimation of ${f}_{i_1,i_2,\theta}$ can be done similarly as in \citet{hormann2015dynamic}. Specifically, we first estimate $C_{i_1,i_2,g}(t,s)$ by
\begin{equation}\label{CCov_est}
    \hat{C}_{i_1,i_2,g}(t,s)=\frac{1}{J}\sum_{j=1}^{J-g}{\varepsilon}_{i_1(j+g)}(t){\varepsilon}_{i_2j}(s),
\end{equation}
$\forall$ $i_1,i_2\in V$, $g\in \mathbb{Z}$, and $t,s\in [0,1]$.
Then, we use a lag-window estimator with the Bartlett window \citep{brillinger2001time} to estimate $f_{i_1,i_2}(t,s|\theta)$ as
\begin{equation}\label{lwee}
\hat{f}_{i_1,i_2}(t,s|\theta)=\frac{1}{2\pi}\sum_{|g|\leq r}\bigg(1- \frac{|g|}{r}\bigg)\hat{C}_{i_1,i_2,g}(t,s)\exp(\operatorname{i}g\theta),
\end{equation}
where $r$ is a bandwidth. The value of $r$ affects the theoretical properties of the resulting estimator, which we will discuss in Section \ref{aps}.

Next, note that for each $\theta\in [-\pi,\pi]$, $\sum_{i=1}^p{f}_{i,i}(t,s|\theta)$ admits a spectral decomposition: $ \sum_{k=1}^{\infty}\nu_k(\theta)\overline{\psi_{k}(t|\theta)} {\psi_{k}(s|\theta)}$ under dynamic weak separability \eqref{ws},
where $\nu_k(\theta)=\operatorname{tr}\big\{\bm{\eta_k}(\theta)\big\}$. 
For any given $\theta\in[-\pi,\pi]$, we estimate ${\psi}_k(\cdot|\theta)$ and $\nu_k(\theta)$ by conducting spectral decomposition for $\sum_{i=1}^p\hat{f}_{i,i}(t,s|\theta)$. 
Denote $\hat{\psi}_k(\cdot|\theta)$ and $\hat{\nu}_k(\theta)$ as the corresponding estimators. Then based on (\ref{dfpca}), we estimate ${\phi}_{kl}(\cdot)$ as
\begin{eqnarray*}
    \hat{\phi}_{kl}(\cdot):=\frac{1}{2\pi}\int_{-\pi}^{\pi}\hat{\psi}_{k}(\cdot|\theta)\exp(-\operatorname{i}l\theta) \ \mathrm{d}\theta.
\end{eqnarray*}
Different from \citet{hormann2015dynamic}, we here pool the spectral densities $\hat{f}_{i,i,\theta}$s for all $i\in V$ to estimate a comment set of functional filters, rather than estimating them separately for each $i$. This enhances the statistical efficiency of functional filters as shown in Section \ref{aps}.

In practice, $\{\bm{\varepsilon}_j;j=1,\cdots,J\}$ cannot be observed directly, and we may apply methods by pooling data information across functions \citep[e.g.,][]{yao2005functional} to estimate the above quantities. However,  these methods may require intensive computations for estimating $\{f_{i_1,i_2,\theta};i_1,i_2\in V, \theta\in [-\pi,\pi]\}$. Considering this, we conduct a pre-smoothing for the MFTS data by assuming the functional data is densely observed. With this, we first obtain estimates of $X_{ij}(\cdot)$ and $\sigma^2_i$, denoted as $\hat{X}_{ij}(\cdot)$ and $\hat{\sigma}_i^2$. Subsequently, we estimate $\mu_i(\cdot)$ by $\hat{\mu}_i(\cdot)=\frac{1}{J}\sum_{j=1}^J\hat{X}_{ij}(\cdot)$, and then estimate $f_{i_1,i_2,\theta}$, $\psi_k(\cdot)$ and $\phi_{kl}(\cdot)$ by approximating $\varepsilon_{ij}(\cdot)$ as $\hat{X}_{ij}(\cdot)-\hat{\mu}_i(\cdot)$ in \eqref{CCov_est}.
In Part B.1 of Supplementary Materials, we present the construction of the above estimators based on the pre-smoothed functional data. 

Furthermore, we estimate the $(i_1,i_2)^{th}$ element of the eigen-matrix $\bm{\eta}_{k}(\theta)$ by
\begin{equation}\label{eigen-matrix_est}
    \hat{\eta}_{i_1,i_2,k}(\theta):=\int_0^1\int_0^1 \hat{f}_{i_1,i_2}(t,s|\theta)\hat{\psi}_{k}(t|\theta)\overline{\hat{\psi}_{k}(s|\theta)}\ \mathrm{d}t\mathrm{d}s.
\end{equation}
After that, we might estimate ${\bm{\Phi}}_{k}(\theta)$ as $\left\{\hat{\bm{\eta}}_k(\theta)\right\}^{-1}$, $\forall \theta\in [-\pi,\pi]$.
In our numerical analysis, however, we have discovered that the inverse of $\hat{\bm{\eta}}_k(\theta)$ can be highly unstable when $p>J$. For a more robust estimator, we apply Theorem \ref{th2} to estimate ${\bm{\Phi}}_{k}(\theta)$ by borrowing strength across the matrices $\left\{\bm{\Phi}_k(\theta);\theta\in S\right\}$, where $S$ is a finite set contained in $[-\pi,\pi]$. Let $\bm{\Phi}_k$ be $\left\{\bm{\Phi}_k(\theta);\theta\in S\right\}$.
Based on the partial correlation graph of MFTS, we assume that there exist some pairs of indexes $(i_1,i_2)$ such that $[\bm{\Phi}_k(\theta)]_{i_1,i_2}=0$ for all $\theta\in S$, and propose a joint graphical Lasso estimator \citep{danaher2014joint,jung2015graphical} for $\bm{{\Phi}}_k$ by minimizing
\begin{equation}\label{glp}
     \text{gLasso}(\bm{{\Phi}}_k)
     =\sum_{\theta\in S}\bigg[\operatorname{tr}\big\{ 
     \hat{\bm{\eta}}_k(\theta){\bm{\Phi}}_k(\theta)\big\}-\operatorname{logdet}\big\{{\bm{\Phi}}_k(\theta)\big\}\bigg]+\lambda_k\sum_{i_1\neq i_2}\sqrt{\sum_{\theta \in S}\left|\left[{\bm{\Phi}}_{k}(\theta)\right]_{i_1,i_2}\right|^2},
\end{equation}
where $\operatorname{logdet}(\cdot)$ denotes the logarithmic determinant of a matrix, and $\lambda_k>0$ is a regularization parameter. 
We denote the above estimator for $\bm{\Phi}_k(\theta)$ as ${\bm{\hat\Phi}}_{k,\lambda_{k}}(\theta)$.
When $\lambda_k=0$, ${\bm{\hat\Phi}}_{k,\lambda_{k}}(\theta)$ reduces to the inverse of $\hat{\bm{\eta}}_k(\theta)$. As $\lambda_k$ increases, the sparsity would be imposed to the grouped terms in \eqref{glp}. The selection of $\lambda_k$ will be discussed in Section \ref{aps}.
Following \citet{danaher2014joint}, we use the alternating direction method of multipliers \citep[ADMM;][]{boyd2011distributed} to solve the above minimization problem. 

\subsection{Conditional Expectation for Score Extraction}\label{cese}

Now we focus on estimating $\bxi_{\cdot,jk}$.
In \citet{hormann2015dynamic}, the scores are estimated using integration (\ref{dfpca2}), where $\varepsilon_{ij}$s for $j=1,\ldots,J$ are approximated by their pre-smoothing, while $\varepsilon_{ij}$ for $j>J$ or $j<1$ are assumed to be $0$. This approach gives rise to biases for the scores at the boundaries, as it relies on the unobserved functional time series outside the time period. Moreover, the integration method also ignores graphical interactions of the MFTS, resulting in a loss of statistical efficiency for score extraction.
To avoid these issues, one may calculate the conditional expectation of the scores given $\bm{Y}$, similar to \citet{yao2005functional}.
Nevertheless, such an approach requires inverting the covariance matrix of $\bm{Y}$, whose dimension is $pJZ\times pJZ$, and is computationally challenging when $pJZ$ is large.

In this subsection, we propose a computationally more efficient approach to calculate the conditional expectation and to hence estimate the scores. To that end, we assume that the multivariate Gaussian process $\bvarepsilon_{j}$ follows
\begin{equation}\label{dy_model}
    \bvarepsilon_{j}(t)=\sum_{k=1}^{{K}}\sum_{|l|\leq {L}_k}\phi_{kl}(t)\bxi_{\cdot,(j+l)k},
\end{equation}
where ${K}$ and ${L}_k$ are fixed and their values can be selected according to Theorem \ref{p2} and equation \eqref{Prop_1}; see Part B.2 in Supplementary Materials for more details. 
We define $\bm{\Sigma}:=\text{diag}({\sigma}_1^2,\cdots,{\sigma}_p^2)$, $\bxi_{k}:=\big(\bxi_{\cdot, (1-L_k)k},\cdots,\bxi_{\cdot, (J+L_k)k}\big)$, and use $\bm{\mu}$, $\bm{\phi}$ and $\bm{\Phi}$ to denote $\big\{\mu_i(\cdot);i\in V\big\}$, $\big\{\phi_{kl}(\cdot);l\leq L_k,k\leq K\big\}$ and $\big\{\bm{\Phi}_k;k\leq K\big\}$, where $\bm{\Phi}_k$ denotes $\left\{\bm{\Phi}_k(\theta);\theta\in [-\pi,\pi]\right\}$. Let $C$ be a constant that may take different values but is nevertheless unrelated to the scores. By \eqref{model_dat} and \eqref{dy_model}, the log-conditional density of the scores is
\begin{equation}\label{pol}
f_{d}\left(\bxi_{1},\cdots,\bxi_{K} |\bm{Y},\bm{\Sigma},\bm{\mu},\bm{\phi},\bm{\Phi}\right)
= f_{d}(\bm{Y}| \bxi_{1},\cdots,\bxi_{K}, \bm{\Sigma},\bm{\mu},\bm{\phi})+ \sum_{k=1}^Kf_{d}(\bxi_{k}| \bm{\Phi}_k)+C,
\end{equation}
where $f_{d}(\bm{Y}| \bxi_{1},\cdots,\bxi_{K}, \bm{\Sigma},\bm{\mu},\bm{\phi})=-\sum_{i=1}^p\sum_{j=1}^J\sum_{z=1}^{Z}\big\{Y_{ijz}-{\mu}_i(t_{z})-\sum_{k\leq K}\sum_{|l|\leq L_k}{\phi}_{kl}(t_{z})$ $\xi_{i(j+l)k}\big\}^2/ (2\sigma^2_i)$ and $f_{d}(\bxi_{k}| \bm{\Phi}_k)$ is the log-marginal density of the stationary multivariate time series $\bxi_{k}$ determined by $\bm{\Phi}_k$. Here, the parameters $\bm{\Sigma}$, $\bm{\mu}$, $\bm{\phi}$ and $\bm{\Phi}$ are taken as their 
estimates given in Section \ref{est of ef}.
Under the Gaussian assumption, the maximizers of $\bxi_{1},\cdots,\bxi_{K}$ in \eqref{pol} can approximate the conditional expectation of the scores given the taken values of parameters. 

To maximize \eqref{pol}, we need to invert the covariance matrix of $\bm{\xi}_{k}$ in $f_{d}(\bm{\xi}_{k}| \bm{\Phi}_k)$, whose dimension is $p(J+2L_k)\times p(J+2L_k)$, for $k=1,\ldots,K$. When $p(J+2L_k)$ is large, this will require large computer memory to store these matrices and high computational costs to invert them. Alternatively, we apply Whittle likelihood \citep{whittle1961gaussian} to construct a computationally tractable pseudo-likelihood of scores. To that end, define $\tilde{\bxi}_{k}(\theta_j):=\bxi_k\bm{\rho}_k(\theta_j)$, where $\bm{\rho}_k(\theta_j)=\frac{1}{\sqrt{2\pi (J+2L_k)}}$ $\big(\exp(-\operatorname{i}1\theta_j),\cdots,\exp(-\operatorname{i}(J+2L_k)\theta_j)\big)^T$ and $\theta_j:=\frac{2\pi j}{J}$, $1\leq j\leq J$.
Under some regularity conditions,  $\tilde{\bxi}_{k}(\theta_j)$, $j=1,\cdots,J$, is approximately a zero-mean complex Gaussian random vector with the covariance matrix $\bm{\eta}_{k}(\theta_j)$, and $\tilde{\bxi}_{k}(\theta_{j_1})$ and $\tilde{\bxi}_{k}(\theta_{j_2})$ are independent with each other if $j_1\ne j_2$ \citep{brillinger2001time}. 
Accordingly, the log-Whittle likelihood \citep{dunsmuir1979central} for $\bxi_{k}$ is given as
\begin{equation}\label{WL}
\tilde{f}_{d}(\bxi_{k}|{\bm{\Phi}}_{k})=
-\frac{1}{2}\sum_{j=1}^{J}\left[\{\tilde{\bxi}_{k}(\theta_j)\}^*\bm{\Phi}_k(\theta_j)\tilde{\bxi}_{k}(\theta_j)-\operatorname{logdet}\left\{{\bm{\Phi}}_k(\theta_j)\right\}\right]
+C.
\end{equation}
Note that we just need to deal with $J$ $p\times p$ matrices in $\tilde{f}_d(\bm{\xi}_{k}| \bm{\Phi}_k)$ with minor changes in statistical properties \citep{dunsmuir1979central}.
As such, we estimate $\xi_{ijk}$ via maximizing the conditional density \eqref{pol} by replacing $f_{d}(\bxi_{k}|{\bm{\Phi}}_{k})$ with 
$\tilde{f}_d(\bxi_{k}|{\bm{\Phi}}_{k})$. We propose a gradient ascend algorithm to iteratively find the maximum; more information about this procedure can be found in Part B.3 in Supplementary Materials.
Given $\hat{\bm{\mu}}$, $\hat{\bxi}_{1},\cdots,\hat{\bxi}_{K}$ and $\bm{\hat\phi}$, we reconstruct $X_{ij}(t)$ by
the truncated representation
$\hat{X}_{ij}(t):=\hat{\mu}_i(t)+\sum_{k\leq K}\sum_{|l|\leq L_k}\hat{\phi}_{kl}(t)\hat{\xi}_{i(j+l)k}$. 

Note that the above procedures are developed for GDFPCA. Based on Theorem \ref{th33}, we may instead model the MFTS by GSFPCA. For this, we replace the log-density of $\bm{Y}$ in \eqref{pol} by $-\sum_{i=1}^p\sum_{j=1}^J\sum_{z=1}^{Z}\big\{Y_{ijz}-\hat{\mu}_i(t_{z})-\sum_{k\leq K}\hat{\varphi}_{k}(t_{z})\xi_{ijk}\big\}^2 / (2\hat{\sigma}_i^2)$ for score extraction, where $\big\{\hat{\varphi}_{k}(\cdot);k\leq K\big\}$, are estimated by the eigenfunctions of the kernel $\sum_{i=1}^p\hat{C}_{i,i,0}(t,s)$ \citep{zapata2019partial}. After that, we reconstruct $X_{ij}(t)$ by $\hat{\mu}_i(t)+\sum_{k\leq K}\hat{\varphi}_{k}(t)\hat{\xi}_{ijk}$.

\subsection{Statistical Properties}\label{aps}
In this subsection, we investigate statistical properties of the first-step estimation in GDFPCA. We  only focus on the case of fully observed processes $\{\bvarepsilon_j;j=1,\cdots,J\}$. 
We assume a general condition of weak dependence called the $L^4$-$m$-$\text{approximablility}$ \citep{hormann2010weakly} for $\{\bvarepsilon_j;j\in \mathbb{Z}\}$ to establish the consistency of $\hat{f}_{i_1,i_2,\theta}$. This condition leads to the weak stationarity \eqref{wst} and $\sup_{i_1,i_2\in V}\sum_{g\in \mathbb{Z}}||C_{i_1,i_2,g}||<\infty$ in Section \ref{me}; see Part A.6 in Supplementary Materials for the definition of $L^4$-$m$-$\text{approximablility}$. 

\begin{theo}
\label{Th3}
Assuming that $\{\bvarepsilon_{j};j\in \mathbb{Z}\}$ is $L^4$-$m$-$\text{approximable}$, we have
\begin{equation*}
\mathbb{E}\sup_{\theta\in[-\pi,\pi]}\left\|{f}_{i_1,i_2,\theta}-\hat{f}_{i_1,i_2,\theta}\right\|=O\bigg(\frac{r}{\sqrt{J}}+\frac{1}{r} \sum_{|g|\leq r}|g|\cdot \big\|C_{i_1,i_2,g}\big\|+\sum_{|g|>r}\big\|C_{i_1,i_2,g}\big\|\bigg),
\end{equation*}
for $i_1,i_2\in V$,
as $J,r\rightarrow\infty$ with $r=o(J^{1/2})$.
\end{theo}
In Theorem \ref{Th3}, the value of $r$ affects the convergence rate of $\hat{f}_{i_1,i_2,\theta}$, and there exists a trade-off for the value of $r$ balancing the different error terms. 
We simply set $r=J^{0.4}$ similar to \citet{hormann2015dynamic}, so that $\hat{f}_{i_1,i_2,\theta}$ converges to ${f}_{i_1,i_2,\theta}$ as $J\rightarrow \infty$. 

Since $\psi_k(\cdot|\theta)$ and $\hat{\psi}_k(\cdot|\theta)$ can only be identified up to some multiplicative factor on the complex unit circle, we need to add an identifiability condition for $\hat{\psi}_k(\cdot|\theta)$ to examine the consistency of functional filters. 
For this purpose, we always adjust $\hat{\psi}_{k}(\cdot|\theta)$ s.t.
$\langle\psi_{k}(\cdot|\theta),\hat{\psi}_{k}(\cdot|\theta)\rangle\geq  0$ for a given $\psi_k(\cdot|\theta)$. This can be achieved without loss of generality, by replacing $\hat{\psi}_{k}(\cdot|\theta)$ as $\hat{\psi}_{k}(\cdot|\theta)\cdot \frac{|\langle\psi_{k}(\cdot|\theta),\hat{\psi}_{k}(\cdot|\theta)\rangle|}{\langle\psi_{k}(\cdot|\theta),\hat{\psi}_{k}(\cdot|\theta)\rangle}$ when $\langle\psi_{k}(\cdot|\theta),\hat{\psi}_{k}(\cdot|\theta)\rangle\neq 0$.

\begin{theo}\label{thh}
Under the $L^4$-$m$-$\text{approximablility}$ of $\{\bvarepsilon_{j};j\in \mathbb{Z}\}$ and  the dynamic weak separability (\ref{ws}), we further assume for $k,k^{\prime}\leq K$, 
\begin{equation}\label{ass1}
\inf_{\theta\in[-\pi,\pi],k^{\prime}\neq k} \left|\frac{1}{p}\sum_{i=1}^p\bigg\{\eta_{i,i,k^{\prime}}(\theta)-{\eta_{i,i,k}}(\theta)\bigg\}\right|> 0,
\end{equation}
where $K$ is finite. Then for $k\leq K$,
\begin{eqnarray*}
    \sup _{l \in \mathbb{Z}}\left\|\phi_{kl}-\hat{\phi}_{kl}\right\|=O_{\operatorname{p}}\bigg[\bigg(\frac{1}{\sqrt{p}}+ \varrho_{p,J,r}\bigg)\cdot\bigg\{\frac{r}{\sqrt{J}}+\sup_{i\in V}\bigg(\frac{1}{r} \sum_{|g|\leq r}|g|\cdot \big\|C_{i,i,g}\big\|+\sum_{|g|>r}\big\|C_{i,i,g}\big\|\bigg)\bigg\}\bigg],
\end{eqnarray*}
as $J,r\rightarrow\infty$ with $r=o(J^{1/2})$, 
where $\varrho_{p,J,r}={{\mathbb{E}(\sup_{{1\leq i_1\neq i_2\leq p}}\mathcal{C}_{i_1,i_2}})/ {{\mathbb{E}(\sup_{i\in V}\mathcal{C}_{i,i} }}})$ with  $\mathcal{C}_{i_1,i_2}=\sup_{\theta\in [-\pi,\pi]}\sqrt{\|{f}_{i_1,i_1,\theta}-\hat{f}_{i_1,i_1,\theta}\|\cdot \|{f}_{i_2,i_2,\theta}-\hat{f}_{i_2,i_2,\theta}\|}$, and $p$ can be finite or tends to $\infty$.
\end{theo}

In Theorem \ref{thh}, condition \eqref{ass1} assumes that the average eigenvalue gap across different individuals is bounded away from zero, introducing the identifiability for the corresponding eigenfunctions in \eqref{ws}. In the rate of $sup _{l \in \mathbb{Z}}\left\|\phi_{kl}-\hat{\phi}_{kl}\right\|$, ${1/\sqrt{p}}{}$ comes from averaging out information from $p$ individual series of MFTS, and $\varrho_{p,J,r}$ arises from the correlations among these individual series.
 It can be shown that $\varrho_{p,J,r}\leq 1$, indicating that 
$sup _{l \in \mathbb{Z}}\left\|\phi_{kl}-\hat{\phi}_{kl}\right\|$ would not blow up even if $p$ diverges faster than $J$.  Furthermore, when the correlations among the individual series are relatively mild, e.g., $\rho_{p,J,r}=o(1)$ as $p,J,r\rightarrow \infty$, a large $p$ also results in a ``blessing" of high-dimensionality for estimating functional filters.
This demonstrates the superiors of \eqref{ws} for improving statistical efficiency.

For each $\theta$ and $k\leq K$, Theorems \ref{Th3} and \ref{thh} also suggest that $\hat{{\eta}}_{i_1,i_2,k}(\theta)$ in \eqref{eigen-matrix_est} is a consistent estimate of ${{\eta}}_{i_1,i_2,k}(\theta)$, $\forall i_1,i_2\in V$; see Part A.7 of Supplementary Materials for more details. 
Then for $\theta \in S$, ${\bm{\hat\Phi}}_{k,\lambda_{k}}(\theta)$ in \eqref{glp} converges to ${\bm{\Phi}}_{k}(\theta)$ under the latent graph $(V,E_k)$ assumption, where $E_k\subset E$ is the edge set in \eqref{uni_gra} for the $k^{th}$ component; see \citet{jung2015graphical} for the detailed conditions and deviations.
To select a suitable $\lambda_k$ in ${\bm{\hat\Phi}}_{k,\lambda_{k}}(\theta)$ for $\theta\in S=\{\theta_1,\cdots,\theta_J\}$, we use the Akaike information criterion (AIC) of the Whittle likelihood (\ref{WL}) by minimizing 
\begin{eqnarray*}
\text{AIC}(\lambda_k)= \sum_{j=1}^J\bigg[\operatorname{tr}\big[\tilde{\bxi}_{k}(\theta_j)\{\tilde{\bxi}_{k}(\theta_j)\}^*\hat{\bm{\Phi}}_{k,\lambda_k}(\theta_j)\big]-\log\det\big\{{\hat{\bm{\Phi}}}_{k,\lambda_k}(\theta_j)\big\}\bigg] + 2\text{df}(\lambda_k),
\end{eqnarray*}
where $\text{df}(\lambda_k)$ measures the model complexity.
Since $\mathbb{E}\tilde{\bxi}_{k}(\theta_j)\{\tilde{\bxi}_{k}(\theta_j)\}^*\approx {\bm{\eta}}_{k}(\theta_j)$ for large $J$ and $\tilde{\bxi}_{k}(\theta_j)$ is latent, we approximate $\tilde{\bxi}_{k}(\theta_j)\{\tilde{\bxi}_{k}(\theta_j)\}^*$ within the AIC by $\hat{\bm{\eta}}_{k}(\theta_j)$. It is difficult to obtain a closed-form expression for $\text{df}(\lambda_k)$, as $\hat{\bm{\Phi}}_{k,\lambda_k}(\theta)$ is a smooth function of $\theta$ due to the lag-window estimator \eqref{lwee}. In Part B.4 in Supplementary Materials, we propose an intuitive estimate for $\text{df}(\lambda_k)$ by making use of the degree of smoothness of $\big\{\hat{\bm{\Phi}}_{k,\lambda_k}(\theta);\theta\in [0,\pi]\big\}$. Based on our numerical experience, $\big\{{\bm{\hat\Phi}}_{k,\lambda_{k}}(\theta);\theta\in S\big\}$ with $\lambda_k$ selected by the AIC are not overly sensitive to the value of $\text{df}(\lambda_k)$. 

\section{Simulation Study}\label{sim}
\subsection{Simulation Setup and Data Generation}\label{simsetup}
In this section, we compare the performance of our proposed GDFPCA and GSFPCA with other FPCA methods for dimension reduction. For simplicity, we only consider the functional data with $\mu_i(t)= 0$, $\forall t\in [0,1]$. We assume that $\{\bvarepsilon_{j};j=1,\cdots,J\}$ follows
\begin{equation}\label{fgts-trunc}
\bvarepsilon_{j}(t)=\sum_{k\leq K}\sum_{|l|\leq L}w_{l}{\phi}_{kl}(t)\bxi_{\cdot,(j+l)k},
\end{equation}
where $L$ is a non-negative integer, $w_{l}$ is a positive weight,
$\big\{\phi_{kl}(\cdot);k\leq K, |l|\leq L\big\}$ is a collection of Fourier basis functions on $[0,1]$, and $\{\bxi_{\cdot,jk};j\in \mathbb{Z}\}$ is a zero-mean $p$-dimensional stationary time series. We set $w_{l}=w^{\prime}_{l}/\sqrt{\sum_{|l|\leq L}(w^{\prime}_{l})^2}$ with $w^{\prime}_{l}={\exp(-2|l|)}$, so that $\sum_{|l|\leq L}w_{l}^2=1$ and a smaller weight is assigned for $\phi_{kl}(\cdot)$ with a larger $|l|$. When $\bxi_{\cdot,jk}$s are independent for different $k$s, $\{\bvarepsilon_{j};j=1,\cdots,J\}$ follows the dynamic weak separability \eqref{ws}. Furthermore, the MFTS satisfies the weak separability \eqref{wss} when $L=0$. To build serial dependencies, we consider an AR(1) model for $\{\bxi_{\cdot,jk};j=1-L,\cdots,J+L\}$ for each $k$, i.e. $\bxi_{\cdot,(j+1)k}=\rho_k \bxi_{\cdot,jk}+\bm{b}_{\cdot,jk}$, where $\bm{b}_{\cdot,jk}$s are independent across different $j$s and $k$s. In our simulation, we set $K=4$ and $\rho_k=0.8, 0.7, 0.6, 0.5$ for $k=1,\cdots,4$. 

\begin{figure}[h]
\begin{center}
\includegraphics[scale = 0.4]{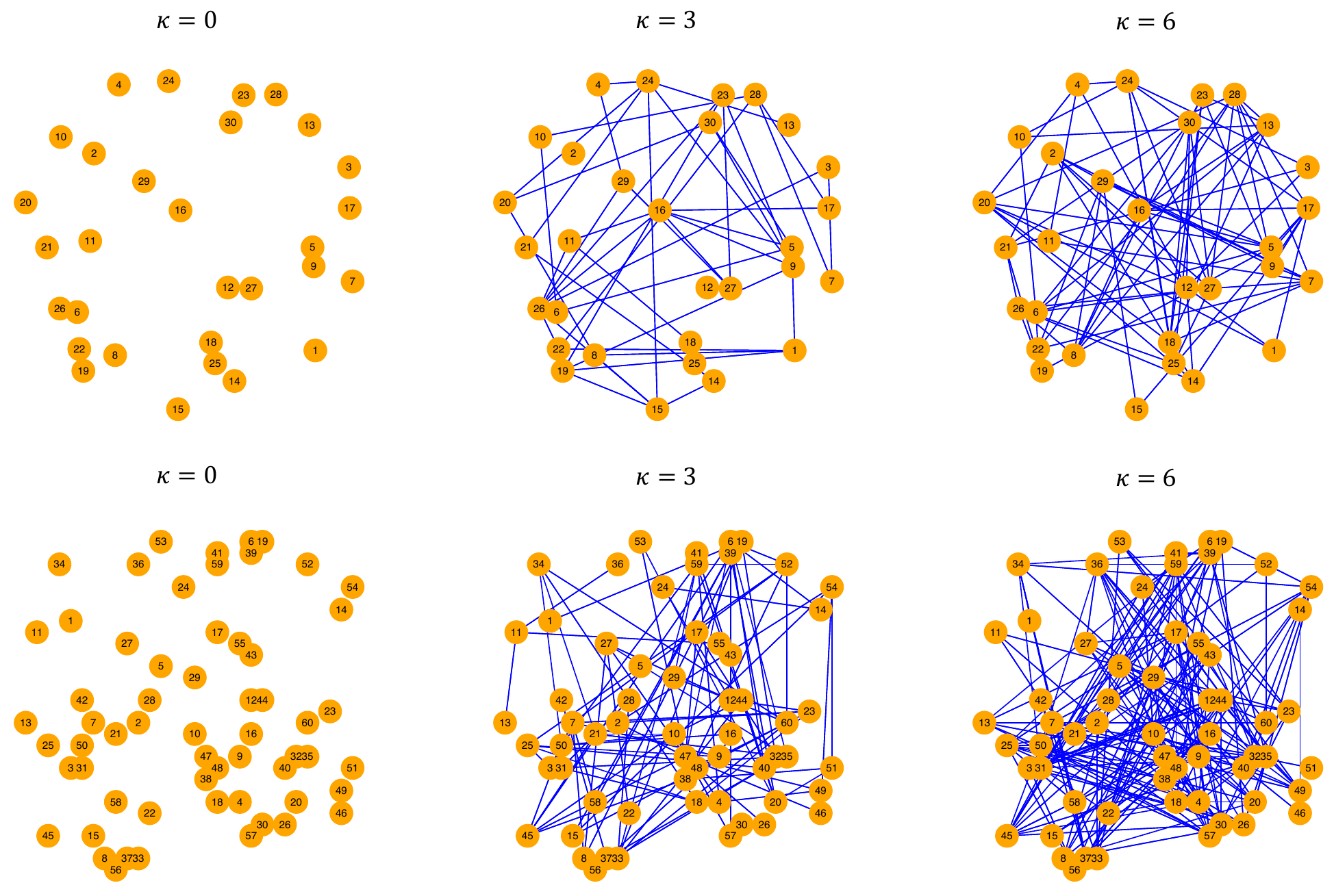}
\end{center}
\caption{Examples of generated graphs with $p=30$ (first row) and $p=60$ (second row).} \label{ngg}
\end{figure}
To add graphical interactions, we first generate an undirected graph. Given the number of nodes $p$, we then randomly generate an edge between any two nodes with probability $\kappa/p$, where $\kappa>0$ controls the sparsity of the edges. Figure \ref{ngg} shows examples of the generated graphs for $\kappa=0,3,6$, corresponding to none, mild and dense connectivities among the nodes. Denote $(V,E)$ as a generated graph. We define $\bm{\Phi}_k^b$ as the precision matrix of the random vector $\bm{b}_{\cdot,jk}$. In particular, we set the $(i_1,i_2)^{th}$ element of $\bm{\Phi}_k^b$ as
\begin{eqnarray*}
    \left[\bm{\Phi}_k^b\right]_{i_1,i_2}=\left\{
\begin{aligned}
&\frac{1}{5}\exp\left(\frac{k}{10}\right)\ \text{if}\ i_1=i_2,\\
&\frac{R_{i_1,i_2}}{5}\exp\left(\frac{k}{10}\right)\ \text{if}\ (i_1,i_2)\in E \ \text{and}\ i_1\neq i_2,\\
&0,\ \text{if}\ (i_1,i_2)\notin E,
\end{aligned}
\right.
\end{eqnarray*}
where $R_{i_1,i_2}\sim\text{Unif}([-0.35, -0.1]\cup[0.1, 0.35])$ controls the partial correlation levels. Accordingly, $\bm{b}_{\cdot,jk}$ is generated from $\operatorname{N}(\bm{0},(\bm{\Phi}_k^b)^{-1})$. By this construction, $\{\bxi_{\cdot,jk};j\in \mathbb{Z}\}$ 
has $(V,E)$ as its partial correlation graph \citep{dahlhaus2003causality}. 

We partition the interval $\mathcal{T}=[0,1]$ into an equally spaced time grid $\{t_z;z=1,\cdots,Z\}$ with $Z=J/4+10$. Under this setting, $Z\rightarrow\infty$ when $J\rightarrow\infty$ so that the functional data is densely observed. We then generate $\varepsilon_{ij}(t_z)$ according to \eqref{fgts-trunc}. For the data contamination, we add an additional noise $\tau_{ijz}$ to $\varepsilon_{ij}(t_z)$, where $\tau_{ijz}$ is independent Gaussian random variables with mean zero and variance {$\sigma^2_i=\mathbb{E}||\varepsilon_{i1}||^2 / 5$ for each given $i$.}

\subsection{Results}\label{simu-result}
We use six different FPCA methods, namely, SFPCA, DFPCA, weakly-separable SFPCA (WSFPCA), weakly-separable DFPCA (WDFPCA), GSFPCA, and GDFPCA, to reconstruct the MFTS $\{\bvarepsilon_{j};j=1,\cdots,J\}$ from the simulated data; see Table \ref{fpca4} for a detailed comparison of these methods. 
Particularly, SFPCA and DFPCA are the univariate FPCAs that reconstruct each individual series of the MFTS separately;
see \citet{ramsay1997functional} and \citet{hormann2015dynamic} for details.
WSFPCA and WDFPCA are the FPCAs based on the weakly-separable KL expansions \eqref{Sta_wes} and \eqref{fgts_model}, respectively.
WSFPCA and WDFPCA estimate the scores as in their univariate versions, except that they combine the information from multiple individual series to estimate the eigenfunctions and functional filters; see \citet{zapata2019partial} and Section \ref{est of ef} for more details.

\begin{table}[h]
\caption{\label{fpca4}Six types of FPCAs.}
\centering
\renewcommand{\arraystretch}{1.5}
\setlength\tabcolsep{0.1pt}
\footnotesize
\begin{tabular}{cccc}
  \hline
 & Representation & Literature  & Scores \\ 
  \hline
SFPCA &  $\varepsilon_{ij}(t)=\sum_{k=1}^{K_i}\hat{\varphi}_{ik}(t)\xi_{ijk}$ & \cite{ramsay1997functional}  & $\hat{\xi}_{ijk}=\langle{\varepsilon}_{ij},\hat{\varphi}_{ik}\rangle$ \\ 
\hline
WSFPCA &  $\varepsilon_{ij}(t)=\sum_{k=1}^K\hat{\varphi}_{k}(t)\xi_{ijk}$ &  \cite{zapata2019partial} & $\hat{\xi}_{ijk}=\langle{\varepsilon}_{ij},\hat{\varphi}_{k}\rangle$\\ 
\hline
GSFPCA &  $\varepsilon_{ij}(t)=\sum_{k=1}^K\hat{\varphi}_{k}(t)\xi_{ijk}$ & --- & Optimization   \\
\hline
DFPCA & $\varepsilon_{ij}(t)=\sum_{k\leq K_i}\sum_{|l|\leq L_{k,i}}\hat{\phi}_{ikl}(t)\xi_{i(j+l)k}$ & \cite{hormann2015dynamic} & $\hat{\xi}_{ijk}=\sum_{|l|\leq L_{k,i}}\langle \varepsilon_{i(j-l)}, \hat{\phi}_{ikl}\rangle$\\
\hline
WDFPCA &  $\varepsilon_{ij}(t)=\sum_{k\leq K}\sum_{|l|\leq L_k}\hat{\phi}_{kl}(t)\xi_{i(j+l)k}$ & --- & $\hat{\xi}_{ijk}=\sum_{l\leq |L_{k}|}\langle \varepsilon_{i(j-l)}, \hat{\phi}_{kl}\rangle$\\
\hline
GDFPCA &  $\varepsilon_{ij}(t)=\sum_{k\leq K}\sum_{|l|\leq L_k}\hat{\phi}_{kl}(t)\xi_{i(j+l)k}$ & --- & Optimization \\ 
   \hline
\end{tabular}
\end{table}

Moreover, GSFPCA and GDFPCA are the graphical versions of WSFPCA and WDFPCA that additionally consider graphical interactions for the score extraction.
For GSFPCA and GDFPCA, we reconstruct the MFTS in the same way as WSFPCA and WDFPCA, except that we extract the scores by optimizing their conditional densities.

In addition to the aforementioned six FPCAs, we also consider two other methods, denoted as (G)SFPCA and (G)DFPCA, which are counterparts to GSFPCA and GDFPCA with the graph being known. Different from GSFPCA and GDFPCA, we set
\begin{eqnarray*}
    \bm{\hat{\Phi}}=\arg\min_{\{{\bm{\Phi}}_k(\theta_j)\in \mathcal{M}_E;k\leq K, j\leq J\}} \sum_{k=1}^K\sum_{f=1}^{J}\bigg
   [\operatorname{tr}\big\{ 
     \hat{\bm{\eta}}_k(\theta_j){\bm{\Phi}}_k(\theta_j)\big\}-\operatorname{logdet}\big\{{\bm{\Phi}}_k(\theta_j)\big\}\bigg],
\end{eqnarray*}
where $\mathcal{M}_E$ denotes the collection of all $p\times p$ positive-definite matrices with the $(i_1,i_2)^{th}$ element being 0 when $(i_1,i_2)\notin E$. This indicates that $ \bm{\hat{\Phi}}$ is estimated by incorporating the known graphical constraints according to Theorem \ref{th2}. We use Algorithm 17.1 in \citet{friedman2001elements} to obtain $\bm{\hat{\Phi}}$ for the score extractions of (G)SFPCA and (G)DFPCA.

We conduct 100 simulations for each setting. To compare the reconstruction accuracy, we define the normalized mean square error $\text{NMSE}(q)$ using the first $q$ components:
\begin{eqnarray*}
\text{NMSE}(q)=\frac{\sum_{i=1}^p\sum_{j=1}^J||\varepsilon_{ij}-\hat{\varepsilon}_{ij}^{q}||^2}{\sum_{i=1}^p\sum_{j=1}^J||\varepsilon_{ij}||^2}\cdot 100\%,
\end{eqnarray*}
where $\hat{\varepsilon}_{ij}^{q}(t)$ for WDFPCA and GDFPCA is defined as $\sum_{k=1}^{\min({K}, q)}\sum_{|l|\leq {L}_k} \hat{\phi}_{kl}(t) \hat\xi_{i(j+l)k}$, with ${K}$ and ${L}_k$ selected by the ratio of variance explained and the cumulative norm, respectively; see part B.2 in Supplementary Materials for their definitions.
For other static FPCAs, $\hat{\varepsilon}_{ij}^{q}(\cdot)$ are defined
analogously according to Table \ref{fpca4}.

\begin{table}[ht!]
\caption{\label{comp} The $\text{NMSE}(q)$ of different FPCAs when the dynamic weak separability \eqref{ws} is achieved. In each setting, we highlight the best performance of the FPCAs in bold (excluding (G)SFPCA and (G)DFPCA).}
\centering
\renewcommand{\arraystretch}{1.4}
\setlength\tabcolsep{1pt}
\footnotesize
\begin{tabular}{c|c|c|cccc|cccc|cccc}
  \hline
\multicolumn{3}{c|}{ \multirow{2}{*}{$\text{NMSE}(q)\ (\%)$}}
   &  \multicolumn{4}{c|}{$\kappa=0$} &  \multicolumn{4}{c|}{$\kappa=3$}& \multicolumn{4}{c}{$\kappa=6$}\\
 \multicolumn{3}{c|}{}   & $q=1$ & $q=2$ & $q=3$ & $q=4$ & $q=1$ & $q=2$ & $q=3$ & $q=4$ 
  & $q=1$ & $q=2$ & $q=3$ & $q=4$\\ 
  \hline
  \multirow{16}{*}{$p=30$} &
  \multirow{8}{*}{$J=20$} &
  SFPCA & 52.36 & 39.90 & 35.49 & 33.96 & 52.15 & 39.50 & 34.80 & 33.19 & 53.88 & 41.11 & 36.58 & 34.99 \\ 
   &  & WSFPCA & 64.93 & 45.35 & 30.74 & 21.50 & 64.02 & 44.30 & 29.81 & 20.52 & 63.58 & 43.71 & 30.54 & 22.47 \\ 
   &   & GSFPCA & 64.59 & 44.53 & 29.25 & 18.85 & 63.69 & 43.53 & 28.37 & 18.11 & 63.10 & 42.70 & 28.76 & 19.59 \\ 
    &   & (G)SFPCA & 64.59 & 44.53 & 29.25 & 18.84 & 63.67 & 43.48 & 28.31 & 18.06 & 62.98 & 42.48 & 28.47 & 19.28 \\ 
      \cline{3-15}
    &  & DFPCA & \textbf{50.26} & \textbf{39.31} & 37.05 & 36.76 & \textbf{49.99} & \textbf{38.50} & 36.00 & 35.66 & \textbf{51.39} & 40.06 & 37.75 & 37.46 \\ 
      &   & WDFPCA & 63.23 & 42.99 & 29.05 & 19.84 & 61.86 & 41.66 & 27.65 & 19.23 & 60.78 & 40.63 & 28.92 & 22.53 \\ 
      &    & GDFPCA & 63.03 & 41.81 & \textbf{25.93} & \textbf{15.06} & 61.65 & 40.16 & \textbf{24.40} & \textbf{14.13} & 59.94 & \textbf{38.17} & \textbf{24.28} & \textbf{15.82} \\ 
      &     & (G)DFPCA & 63.04 & 41.81 & 25.94 & 15.05 & 61.63 & 40.10 & 24.34 & 14.08 & 59.83 & 38.00 & 24.02 & 15.60 \\ 
  \cline{2-15}
      &   \multirow{8}{*}{$J=40$}&
      SFPCA & 55.63 & 39.00 & 30.41 & 26.06 & 55.73 & 38.93 & 30.26 & 25.77 & 56.33 & 39.32 & 30.76 & 26.41 \\ 
    & & WSFPCA & 63.92 & 42.91 & 27.53 & 17.15 & 63.32 & 42.37 & 27.22 & 16.89 & 62.56 & 41.89 & 27.07 & 17.50 \\ 
   &   & GSFPCA & 63.72 & 42.47 & 26.75 & 15.93 & 63.14 & 41.96 & 26.46 & 15.73 & 62.33 & 41.39 & 26.20 & 16.17 \\ 
    &   & (G)SFPCA & 63.72 & 42.47 & 26.75 & 15.93 & 63.11 & 41.90 & 26.38 & 15.62 & 62.23 & 41.19 & 25.92 & 15.81 \\  
        \cline{3-15}
    & & DFPCA & \textbf{50.89} & \textbf{34.31} & 28.05 & 26.29 & \textbf{50.85} & \textbf{34.11} & 27.65 & 25.75 & \textbf{51.44} & \textbf{34.61} & 28.26 & 26.45 \\ 
       &  & WDFPCA & 61.81 & 39.47 & 23.18 & 12.11 & 60.70 & 38.58 & 22.82 & 12.47 & 58.78 & 36.77 & 22.46 & 13.84 \\ 
       &   & GDFPCA & 61.69 & 38.70 & \textbf{21.58} & \textbf{9.75} & 60.57 & 37.53 & \textbf{20.82} & \textbf{9.50} & 58.36 & 35.38 & \textbf{19.84} & \textbf{10.09} \\ 
        &   & (G)DFPCA & 61.69 & 38.70 & 21.58 & 9.75 & 60.54 & 37.49 & 20.76 & 9.44 & 58.27 & 35.25 & 19.66 & 9.90 \\
        \cline{2-15}
       \hline
 \multirow{16}{*}{$p=60$} &
  \multirow{8}{*}{$J=20$} &
   SFPCA & 51.94 & 39.36 & 34.91 & 33.37 & 51.71 & 39.34 & 35.00 & 33.51 & 58.71 & 42.70 & 37.07 & 35.00 \\ 
& & WSFPCA & 64.88 & 45.08 & 30.28 & 20.96 & 64.85 & 44.74 & 30.05 & 20.86 & 62.77 & 41.84 & 32.28 & 26.79 \\ 
& & GSFPCA & 64.57 & 44.29 & 28.88 & 18.43 & 64.54 & 43.96 & 28.66 & 18.36 & 62.05 & 40.37 & 30.03 & 23.79 \\ 
& & (G)SFPCA & 64.57 & 44.29 & 28.88 & 18.43 & 64.52 & 43.92 & 28.61 & 18.33 & 61.86 & 40.00 & 29.52 & 23.17 \\ 
  \cline{3-15}
&  & DFPCA & \textbf{50.07} & \textbf{38.79} & 36.43 & 36.14 & \textbf{49.81} & \textbf{38.75} & 36.48 & 36.17 & \textbf{53.15} & 39.23 & 36.41 & 36.03 \\ 
& & WDFPCA & 63.34 & 42.57 & 28.38 & 18.79 & 63.22 & 42.22 & 28.23 & 18.82 & 56.27 & 36.21 & 29.11 & 26.30 \\ 
& & GDFPCA & 63.15 & 41.68 & \textbf{25.53} & \textbf{14.53} & 63.01 & 41.25 & \textbf{25.27} & \textbf{14.48} & 54.55 & \textbf{32.34} & \textbf{23.22} & \textbf{18.68} \\ 
& & (G)DFPCA & 63.15 & 41.68 & 25.53 & 14.53 & 62.99 & 41.23 & 25.22 & 14.45 & 54.49 & 32.20 & 23.04 & 18.47 \\
  \cline{2-15}
      &   \multirow{8}{*}{$J=40$}&
       SFPCA & 55.48 & 38.70 & 30.28 & 25.90 & 55.36 & 38.72 & 30.30 & 25.92 & 56.77 & 40.01 & 31.12 & 26.69 \\ 
& & WSFPCA & 63.62 & 42.67 & 27.36 & 16.97 & 63.47 & 42.45 & 27.25 & 16.97 & 59.19 & 38.6 & 25.46 & 17.46 \\ 
& & GSFPCA & 63.43 & 42.24 & 26.60 & 15.78 & 63.28 & 42.03 & 26.50 & 15.78 & 58.93 & 38.08 & 24.61 & 16.26 \\ 
& & (G)SFPCA & 63.43 & 42.24 & 26.60 & 15.78 & 63.26 & 41.99 & 26.44 & 15.71 & 58.79 & 37.79 & 24.16 & 15.67 \\ 
    \cline{3-15}
   & & DFPCA & \textbf{50.79} & \textbf{33.96} & 27.66 & 25.84 & \textbf{50.80} & \textbf{34.13} & 27.93 & 26.15 & \textbf{51.82} & 34.80 & 28.46 & 26.66 \\ 
& & WDFPCA & 61.70 & 39.37 & 22.81 & 11.44 & 61.46 & 39.08 & 22.80 & 11.72 & 53.66 & 31.71 & 20.64 & 15.11 \\ 
& & GDFPCA & 61.58 & 38.65 & \textbf{21.50} & \textbf{9.50} & 61.34 & 38.31 & \textbf{21.32} & \textbf{9.55} & 51.94 & \textbf{28.60} & \textbf{16.22} & \textbf{9.58} \\ 
& & (G)DFPCA & 61.58 & 38.65 & 21.51 & 9.50 & 61.33 & 38.29 & 21.28 & 9.51 & 51.86 & 28.44 & 16.01 & 9.29 \\
       \hline
\end{tabular}
\end{table}

In Table \ref{comp}, we present results of averaged $\text{NMSE}(q)$ from 100 simulations for the cases of $J=20$ and $J=40$. The case of $J=60$ is also given in Part C.1 in Supplementary Materials. In each simulation, the data are generated with $L=1$, and hence the optimal representation of the MFTS differs from its static representation according to Theorem \ref{th33}. Consequently, the dynamic versions of FPCAs are expected to perform better than their static counterparts; this is indeed the case in our simulations for most of the combinations of $p$, $J$, $\kappa$ and $q$. Besides, for a fixed number of $p$, the NMSEs of WDFPCA and GDFPCA become smaller as the number of time units $J$ increases. This supports our statistical properties discussed in Section \ref{aps}. In addition, for any fixed $p$, $J$ and $\kappa$, the NMSEs of the dynamic version of FPCAs are the smallest when $q=4$, the true truncation number $K$ in \eqref{fgts-trunc}. Recall that when calculating $\hat{\varepsilon}_{ij}^{q}(\cdot)$, we use the first ${\min({K}, q)}$ components. In our simulation, we always find $K=4$. For this reason, we only report results up to $q=4$. 

In Part C.1 of Supplementary Materials, we present the variances of the estimated functional filters by different dynamic FPCAs. We discover that the variances of the estimated functional filters from WDFPCA or GDFPCA are significantly smaller than those from DFPCA. These findings provide supporting evidences for Theorem \ref{thh} and also explain the results in Table \ref{comp}, where the averaged $\text{NMSE}(q)$ values of the DFPCA are significantly larger than those of the other dynamic FPCAs when $q=4$.
Furthermore, we find that GDFPCA always outperforms WDFPCA, and the improvement is more significant when $\kappa=6$ and $J=20$. 
In these cases, the functions across different individuals may be highly correlated, and we need to further account for these graphical interactions for score extractions as there is limited information at the temporal level.

Overall, GDFPCA can borrow strength across different individual series in the graph, thereby reducing estimation errors for both functional filters and scores when $J$ is relatively small.
Moreover, note that the performances of GDFPCA are nearly identical to those of (G)DFPCA, respectively, indicating that our proposed joint graphical Lasso estimator for $\bm{\Phi}_k(\theta)$ works well for our purpose.

\begin{table}[h]
\caption{\label{comp2}The $\text{NMSE}(q)$  for $J=40$ and $q=4$ of different FPCAs when the dynamic weak separability \eqref{ws} is not satisfied (case 1) or degenerates to the weak separability \eqref{wss} (case 2).  We highlight the best performance in each setting in bold.}
\centering
\renewcommand{\arraystretch}{1.5}
\footnotesize
\begin{tabular}{c|c|ccc|ccc}
  \hline
  \multicolumn{2}{c|}{ \multirow{2}{*}{$\text{NMSE}(4)\ (\%)$}}
   &  \multicolumn{3}{c|}{Case 1}& \multicolumn{3}{c}{Case 2}\\
 \multicolumn{2}{c|}{} & \multicolumn{1}{c}{$\kappa=0$} &  \multicolumn{1}{c}{$\kappa=3$}& \multicolumn{1}{c|}{$\kappa=6$}&  \multicolumn{1}{c}{$\kappa=0$} &  \multicolumn{1}{c}{$\kappa=3$}& \multicolumn{1}{c}{$\kappa=6$}\\
  \hline
 \multirow{6}{*}{$p=30$} &
SFPCA & 33.32 & 33.25 & 33.64 &  6.11 & 5.86  & 6.45 \\ 
   & WSFPCA & 24.71 & 24.80 & 24.98  & 5.51 & 5.28 & 5.85  \\ 
      & GSFPCA & 23.62 & 23.72 & 23.74 & \textbf{4.68} & \textbf{4.50} &  \textbf{4.94}   \\ 
        \cline{2-8}
&  DFPCA & 32.83 & 32.69 & 33.35 & 11.24 & 10.77 &11.61 \\ 
        & WDFPCA  & 16.82 & 18.47 & 19.30 & 7.59 & 7.51 & 8.42 \\ 
          & {GDFPCA}  & \textbf{13.82} & \textbf{15.22} & \textbf{15.33}  & {4.84} & {4.67} & {5.16} \\
    \cline{1-8}
         \multirow{6}{*}{$p=60$} &
   SFPCA & 33.36 & 33.87 & 32.17 & 6.00 & 6.05 &  6.43\\ 
    & WSFPCA & 24.63 & 24.91 & 26.76 &  5.39 &  5.41 & 5.85 \\ 
     & GSFPCA  & 23.57 & 23.85 & 26.22  & \textbf{4.59} & \textbf{4.61} &  4.89\\ 
        \cline{2-8}
 & DFPCA & 33.11 & 34.05 & 30.93 &  10.95 &  11.04 &  11.50 \\ 
        & WDFPCA & 17.09 & 17.30 & 22.68 &  7.34 & 7.29 &  8.78 \\ 
          & {GDFPCA} & \textbf{14.05} & \textbf{14.43} & \textbf{17.58}  &  {4.72} & {4.73} & \textbf{4.75}\\ 
         \hline
\end{tabular}
\end{table}

To further investigate the robustness of the dynamic weak separability \eqref{ws}, we conduct additional simulations for the cases that the dynamic weak separability \eqref{ws} is not satisfied (case 1) or degenerates to the weak separability \eqref{wss} (case 2). For case 1, we generate the data by \eqref{fgts-trunc}, except that we multiply a fluctuation $1+5\sin(it/p)$ to $ \phi_{kl}(\cdot)$ for each $i$. For case 2, we simply set $L=0$ in \eqref{fgts-trunc} for the data generation. This time we only report results for $J=40$ and $q=4$. 
When the dynamic weak separability \eqref{ws} is not satisfied, the GDFPCA may not be the optimal choice for the reconstruction of MFTS.
Nonetheless, Table \ref{comp2} (case 1) shows that the reconstruction by GDFPCA still performs competitively, owing to its lower estimation uncertainty by pooling together data from different nodes.

When $L=0$, the dynamic weak separability \eqref{ws} still holds, but it degenerates to the weak separability \eqref{wss}. We investigate this case (case 2 in Table \ref{comp2}) to compare the performance of FPCAs. Results show that the static FPCAs outperform their dynamic counterparts in most cases, which is expected based on Theorem \ref{th33}. Normally, the dynamic representation has higher model complexity than its static counterpart, which may result in a poorer reconstruction under a simpler dependence structure. Nevertheless, unlike the DFPCA and WDFPCA, the results of GDFPCA are satisfactory compared with those of their static counterparts. 
This finding suggests that considering graphical interactions for score extractions may alleviate model complexities raised by the dynamic representation.

\section{Data Illustration}\label{real}
For data analysis, we consider hourly readings of PM2.5 concentration (measured in $\mu\text{g}/\text{m}^{3}$) collected from 24 monitoring stations (in three cities in China) in the winter of 2016, with a total length of 60 days. A square-root transformation is employed to stabilize the variance. Then in our notation, we observe a discrete MFTS with $p=24$, $J=60$, and $Z=24$. Examples of daily curves and locations of monitoring stations are illustrated in Figure \ref{dat}. 

\begin{figure}[h]
\begin{center}
\includegraphics[scale = 0.6]{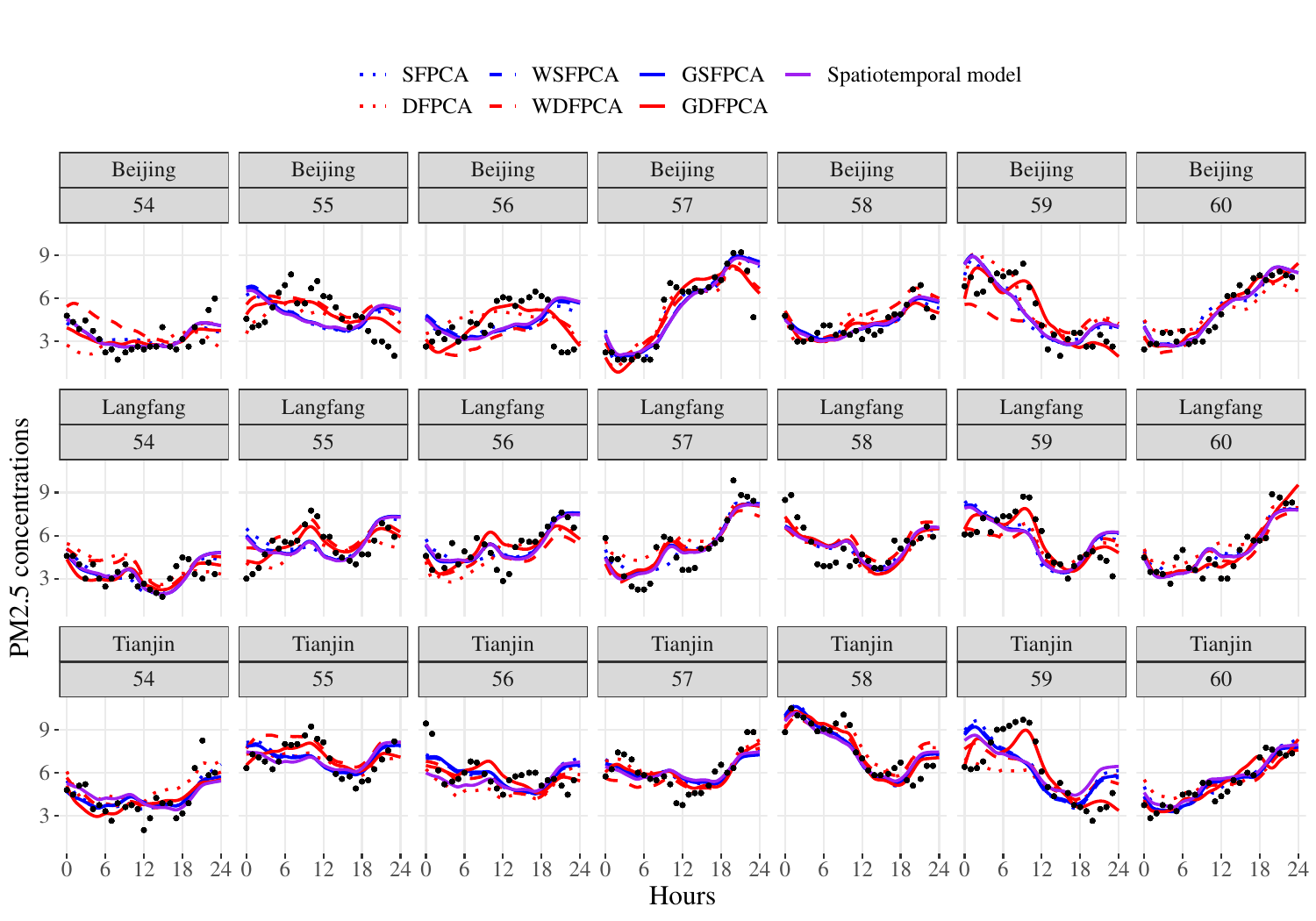}
\end{center}
\caption{Reconstructed curves on the last seven days for three stations in Beijing, Tianjin, and Langfang, respectively. The black dots denote the observed data, whereas the solid and dashed lines denote reconstructed curves using static and dynamic FPCAs and a spatiotemporal model, respectively.} \label{fit_fda}
\end{figure}

We respectively apply the FPCAs in Table \ref{fpca4} to reconstruct the MFTS.
For these FPCAs, we choose their truncation numbers $K_i$ or $K$ in Table \ref{fpca4} using the fractions of variance explained (FVE), with the FVE set to be larger than 80\%; the definitions of FVE can be found in Part B.2 in Supplementary Materials. In our case, $K_i$ takes different values for SFPCA and DFPCA, and $K=2$ for WSFPCA, GSFPCA, WDFPCA and GDFPCA. Besides, $L_{k,i}$ or $L_k$ in Table \ref{fpca4} are determined similarly as in Section \ref{sim}. 
In addition to the FPCA approaches, we also compare a spatiotemporal model similar to the GSFPCA. This model was proposed by \citet{wikle2019spatio}, where they denote $\varepsilon_{ij}(t)$ as $\varepsilon_{j}(\bm{s}_i,t)$, and the spatiotemporal responses are approximated using $\varepsilon_{j}(\bm{s}_i,t)=\hat{\mu}_i(t)+\sum_{k=1}^K\hat{\varphi}_{k}(t)\xi_{jk}(\bm{s}_i)$. Here, $\hat{\mu}_i(\cdot)$, $\hat{\varphi}_{k}(\cdot)$ and $K$ are the same as in GSFPCA, $\bm{s}_i$ is the geographical location of a station, and $\xi_{jk}(\cdot)$ is a zero-mean Gaussian process with stationary exponential covariance function for each pair $(j,k)$. For each $j$, we estimate $\big\{\xi_{jk}(\bm{s}_i);i\in V, k\leq K\big\}$ by their conditional expectations onto $\{Y_{ijz};i\in V, z=1,\cdots,Z\}$ utilizing the geographical information of stations; see Section 4.4.3 of \cite{wikle2019spatio} for more details.

For illustration, we separately pick one station from each city and show the reconstructed MFTS for the last seven days of the study period in Figure \ref{fit_fda}. One can see that the reconstructed MFTS using DFPCA and WDFPCA performs poorly in capturing the data pattern due to their biases for the scores at the boundaries.
Nonetheless, the proposed GDFPCA can better capture the curvature of data. For example, on day 59 in all three cities, there was a temporary rise followed by a big drop of PM2.5 around noon. The reconstructed curves by our method (solid red line) give the best recognition of this pattern among the methods in Figure \ref{fit_fda}.

\begin{figure}
\begin{center}
\includegraphics[scale = 0.5]{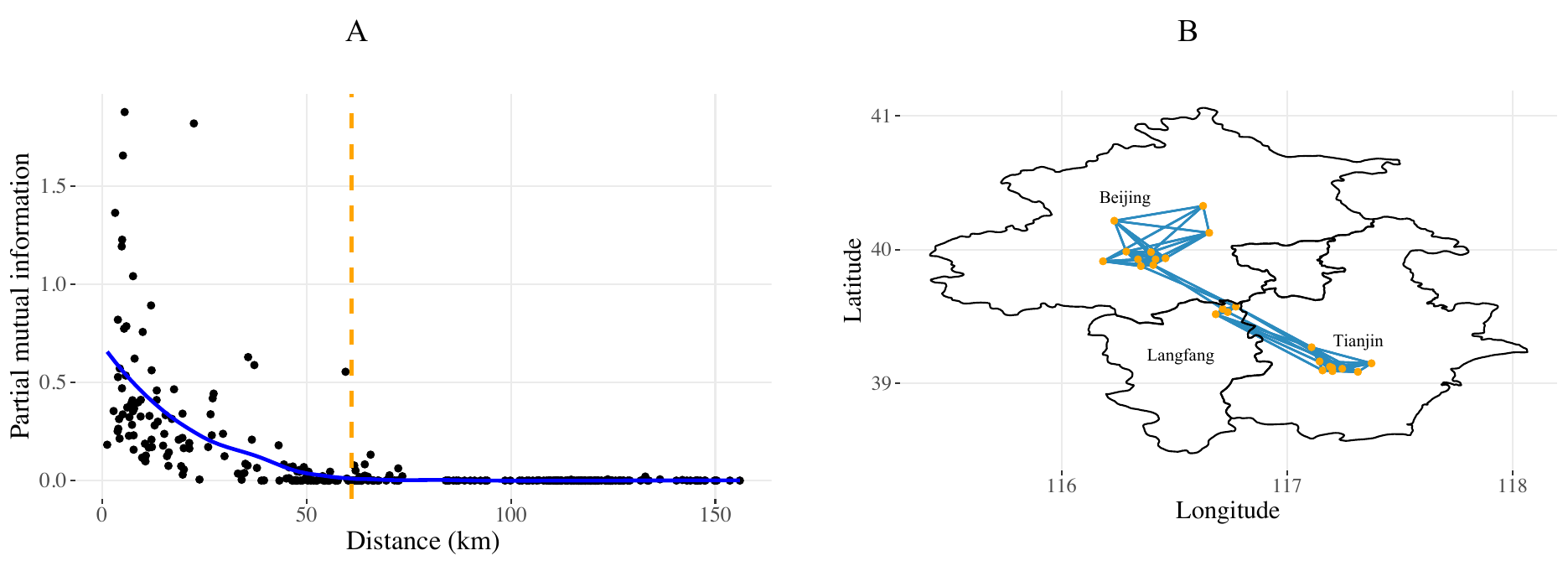}
\end{center}
\caption{\textbf{A}. Scatter plot of partial mutual information versus geographical distances between monitoring stations. The blue line is a local polynomial fit for data points. \textbf{B}. Partial correlation graph for the stations in three cities, where the blue edge between two stations is connected if their partial mutual information is larger than 0.05.} \label{fit_par_cor}
\end{figure}

Unlike the above spatiotemporal model, we do not assume the spatial stationarity condition for the MFTS data; therefore, our proposed GDFPCA is more flexible in capturing spatial dependencies within the temporally correlated functions.
To characterize the spatial dependencies, we utilize the partial mutual information \citep{brillinger1996remarks} for multivariate time series, which is defined as 
\begin{eqnarray*}
    \mathcal{I}_{i_1,i_2} = -\frac{1}{2\pi}\sum_{k=1}^K\sum_{j=1}^J\log\left\{1-\frac{\left|[\hat{\bm{\Phi}}_{k,\lambda_k}(\theta_j)]_{i_1,i_2}\right|^2}{{[\hat{\bm{\Phi}}_{k,\lambda_k}(\theta_j)]_{i_1,i_1}[\hat{\bm{\Phi}}_{k,\lambda_k}(\theta_j)]_{i_2,i_2}}}\right\},
\end{eqnarray*}
where $\hat{\bm{\Phi}}_{k,\lambda_k}(\theta_j)$ is the joint graphical Lasso estimator in Section \ref{est of ef}. In Figure \ref{fit_par_cor} \textbf{A}, we present a scatter plot of the partial mutual information for all pairs of stations versus their geographical distances. It shows that $\mathcal{I}_{i_1,i_2}$ tends to decrease if the distance between stations $i_1$ and $i_2$ gets large. When the distance is larger than {$60$ km ($52.90\%$ of all pairs),  the partial mutual information is almost zero. That means, the connectivity among these stations can be ignored. In Figure \ref{fit_par_cor} \textbf{B}, we present a partial correlation graph for station pairs by calculating their partial mutual information, with a thresholding value for $\mathcal{I}_{i_1,i_2}$ taken as 0.05. As shown in Figure \ref{fit_par_cor} \textbf{B}, we find that the PM2.5 concentration among the stations in Tianjin and Langfang is more partially correlated. }

\begin{figure}[h]
\begin{center}
\includegraphics[scale = 0.5]{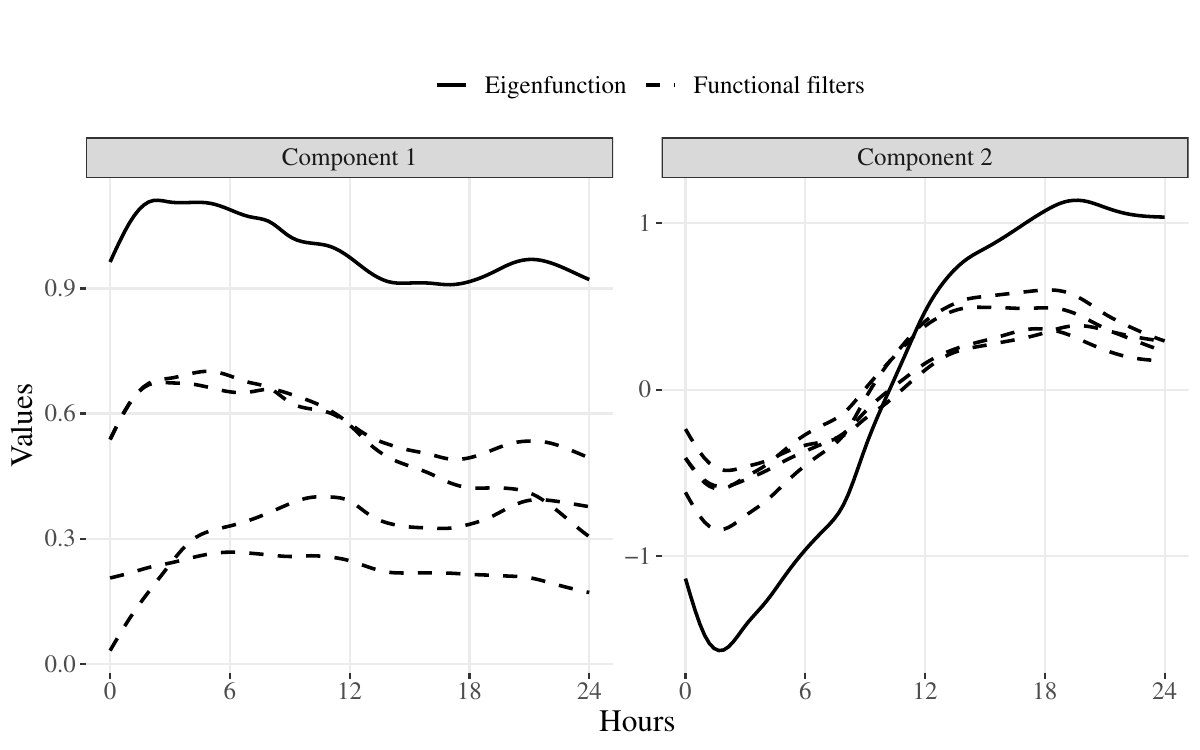}
\end{center}
\caption{The estimated eigenfunction $\hat{\varphi}_k(\cdot)$ and functional filter $\hat{\phi}_{kl}(\cdot)$ for $k=1$ (left panel) and $k=2$ (right panel), respectively. We present the first four $\hat{\phi}_{kl}(\cdot)$s with the largest $||\hat{\phi}_{kl}||$.} \label{mean_eng}
\end{figure}

In Part C.2 in Supplementary Materials, we illustrate the validity of the dynamic weak separability \eqref{ws} for the PM2.5 data.
To further evaluate the static and dynamic representations, we investigate the covariance structures via the estimated eigenfunctions and functional filters. Note that in general, the functional filter $\phi_{kl}(\cdot)$ cannot be uniquely identified, unless the weak separability \eqref{wss} is satisfied. In that case, $\phi_{kl}(\cdot)$, $|l|\le L_{k}$, are proportional to each other for each $k$, and the dynamic FPCA degenerates to its static counterpart according to Theorem \ref{th33}. We compare the eigenfunctions and functional filters of the first two components obtained from GSFPCA and GDFPCA in Figure \ref{mean_eng}.
It shows that the eigenfunctions and functional filters are quite different in shape. This finding suggests that the separability condition \eqref{wss} is not satisfied. Thus, using dynamic FPCAs to reconstruct the underlying MFTS would be more appropriate for the PM2.5 data.

\section{Discussion}\label{disc}
In this study, we develop a theoretical framework to model and reconstruct MFTS from noisy data, considering both serial dependencies and graphical interactions. These two-way dependencies may lead to inefficient dimension reduction when using classical FPCA methods. We propose a key assumption of dynamic weak separability \eqref{ws}, under which we define the partial correlation graph for infinite-dimensional MFTS and facilitate the DFPCA \citep{hormann2015dynamic} to optimally reconstruct signals of MFTS by using graphical-level information. The superior performance of GDFPCA has been demonstrated through a series of simulation studies and a real data example.

Weak separability \citep{lynch2018test,liang2021test} is a novel concept for functional data to characterize the covariance structure among random curves.
According to our theories and simulation study, the form of weak separability plays a prominent role in both defining graphical models and determining optimal approximations. Notably, under the dynamic weak separability condition, we establish a valid partial spectral density kernel to evaluate the partial correlation graph among infinite-dimensional curves. While the theoretical framework in \citet{zapata2019partial} is only applicable for independently sampled curves, we relax this sampling scheme to weakly dependent functional time series. 
Since the scores obtained from the proposed GDFPCA preserve all the graph information, a potential application of our method can be forecasts of functional time series by borrowing strength from other curves in the graph.

To reconstruct the MFTS by the GDFPCA, we follow the joint graphical Lasso method proposed by \citet{danaher2014joint} to estimate $\bm{\Phi}_k(\theta)$, rather than estimating the actual graph $(V,E)$. These are essentially two different topics. Through empirical studies, we have found that the estimated $\bm{\Phi}_{k}(\theta)$ is not sensitive to the reconstruction results, and thus the algorithm proposed in Section \ref{est of ef} is sufficient for dimension reduction. Alternatively, if the study goal is to reveal the actual graph, one may estimate edges using $\cup_{k=1}^{K} E_k$ from \eqref{uni_gra}. To do that, a more sophisticated model selection method may be necessary since the errors from estimating edges $\{E_k, k\leq K\}$ and the selection of the finite truncation $K$ may significantly influence the resulting graph.
In \citet{zapata2019partial}, the unknown graph for independently sampled multivariate functional data was estimated by a modified joint graphical Lasso, where the penalty term was designed to regularize both the sparsity level for each $E_k$ and the common sparsity levels shared by $\cup_{k}^KE_k$. Under our framework, a similar procedure can be potentially developed based on the penalized likelihood \eqref{glp}.

\bibliographystyle{apalike}
\bibliography{refbib}

\end{document}